# Recent Advances in Graphene-Based Humidity Sensors with the Focus of Structural Design: A Review

Hongliang Ma, Jie Ding, Zhe Zhang, Qiang Gao, Quan Liu, Gaohan Wang, Wendong Zhang, Xuge Fan

***Abstract*—The advent of the 5G era means that the concepts of robot, VR/AR, UAV, smart home, smart healthcare based on IoT (Internet of Things) have gradually entered human life. Since then, intelligent life has become the dominant direction of social development. Humidity sensors, as humidity detection tools, not only convey the comfort of human living environment, but also display great significance in the fields of meteorology, medicine, agriculture and industry. Graphene-based materials exhibit tremendous potential in humidity sensing owing to their ultra-high specific surface area and excellent electron mobility under room temperature for application in humidity sensing. This review begins with the introduction of examples of various synthesis strategies of graphene, followed by the device structure and working mechanism of graphene-based humidity sensor. In addition, several different structural design methods of graphene are summarized, demonstrating the structural design of graphene can not only optimize the performance of graphene, but also bring significant advantages in humidity sensing. Finally, key challenges hindering the further development and practical application of high-performance graphene-based humidity sensors are discussed, followed by presenting the future perspectives.**

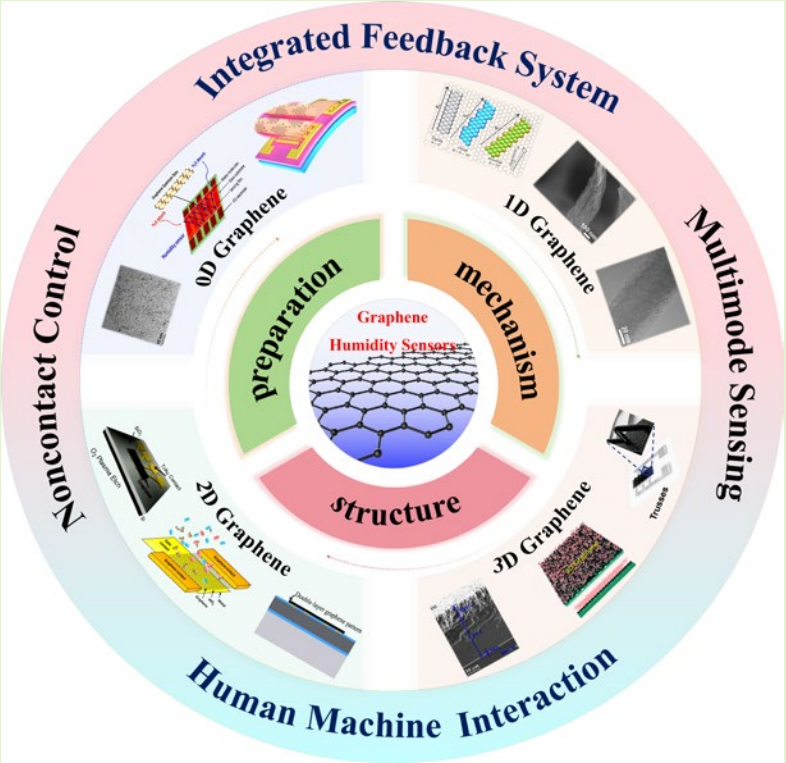

## I. Introduction

HUMIDITY, as an important environmental sensitivity factor in human life, directly conveys the comfort of the living environment and reflects the information of human health [1-6]. Hence, it is imperative to develop advanced humidity sensors with superior detection abilities as the existence of humidity in complex environments affects human health, industrial production [7-10], agricultural production [11-14], and energy supply [15-20]. An ideal high-performance humidity sensor must meet certain basic demands such as high sensitivity, low hysteresis, fast response/recovery time, durability, and low production cost [21-26]. To meet these challenging requirements, many pairs of moisture-sensitive materials have been developed for humidity sensors. Examples include typical ceramics such as alumina [27-31]; two-dimensional (2D) materials that play an important role in humidity sensing due to their large specific surface area and high carrier mobility [32-36], such as transition metal sulfides (TMDS) [37-40], MXene [41-46], black phosphorus (BP) [47-50], and boron nitride (BN) [51, 52]; polymer materials, such as cellulose [53-60], polypyrrole (PPy) [61-64], polyaniline (PANI) [65-68], etc. However, the shortcomings of the above moisture-sensitive materials include the single structure, limited application range, and poor tunability. Thus, developing flexible and structurally tunable moisture-sensitive materials is critically necessary to enhance the versatility of material properties.

Graphene, a monolayer two-dimensional honeycomb lattice structure material formed by carbon atoms connected by $sp^2$ hybridization, exhibits two-dimensional conjugated structure as well as unique energy band structure [69]. The distinctive atomic arrangement of graphene results in a unique electronic structure [70, 71] and extraordinary physicochemical properties [72], providing a wide range of applications in the fields of optoelectronics [73, 74], spintronics [75, 76], energy harvesting [77-79], storage [80], MEMS/NEMS [81-89] and sensors [90, 91]. Moreover, graphene-based materials exhibit great potential in humidity detection due to their ultra-high specific surface area, high electron mobility (15,000 $cm^2/(V\cdot s)$) at room temperature, and extremely high electrical conductivity [92]. Nevertheless, although some outstanding reviews summarized graphene chemistry [93], gases [94, 95] and humidity sensor applications, the application of graphene-based structural design engineering in humidity sensors has not yet been specifically reviewed. In addition, the above reviews lack a detailed summary of the forefront preparation process of graphene and the working mechanism of different humidity sensors. Meanwhile, the majority of the reviews in the application section emphasize the conventional application areas of humidity sensors, such as wearable, non-contact

This work was supported by the the Beijing Natural Science Foundation (Grant No. 4232076), National Natural Science Foundation of China (Grant No. 62171037 and 62088101), National Key Research and Development Program of China (2022YFB3204600), National Science Fund for Excellent Young Scholars (Overseas), Beijing Institute of Technology Teli Young Fellow Program (2021TLQT012), Beijing Institute of Technology Science and Technology Innovation Plan.

Please see the Acknowledgment section of this article for the author affiliations.

sensing, health monitoring, etc., rather than comprehensively expanding on the novel areas.

This review summarizes the research progress of graphene and its derivatives in humidity sensors in recent years, concludes the preparation and properties of graphene in detail, thoroughly explores the structure and mechanism of graphene-based humidity sensors, and especially elucidates the key role of the structural design of graphene in humidity sensors (Fig. 1). Specifically, the first section thoroughly describes eleven recent methods for the preparation of graphene and summarizes the advantages and disadvantages among several different preparation methods. The second section briefly describes seven sensing mechanisms of graphene-based humidity sensors, including resistive, impedance, capacitive, surface acoustic wave (SAW), quartz crystal microbalance (QCM), and field effect (FET), as well as fiber-optic. The third section highlights several different structural designs of graphene and its derivatives, including four different types of structures (0D, 1D, 2D, 3D). Additionally, the progress of the application of graphene with different structures in humidity sensors is reviewed. Finally, the existing challenges, directions, and future trends of graphene-based humidity sensors are elaborated to meet the requirements of future practical applications.

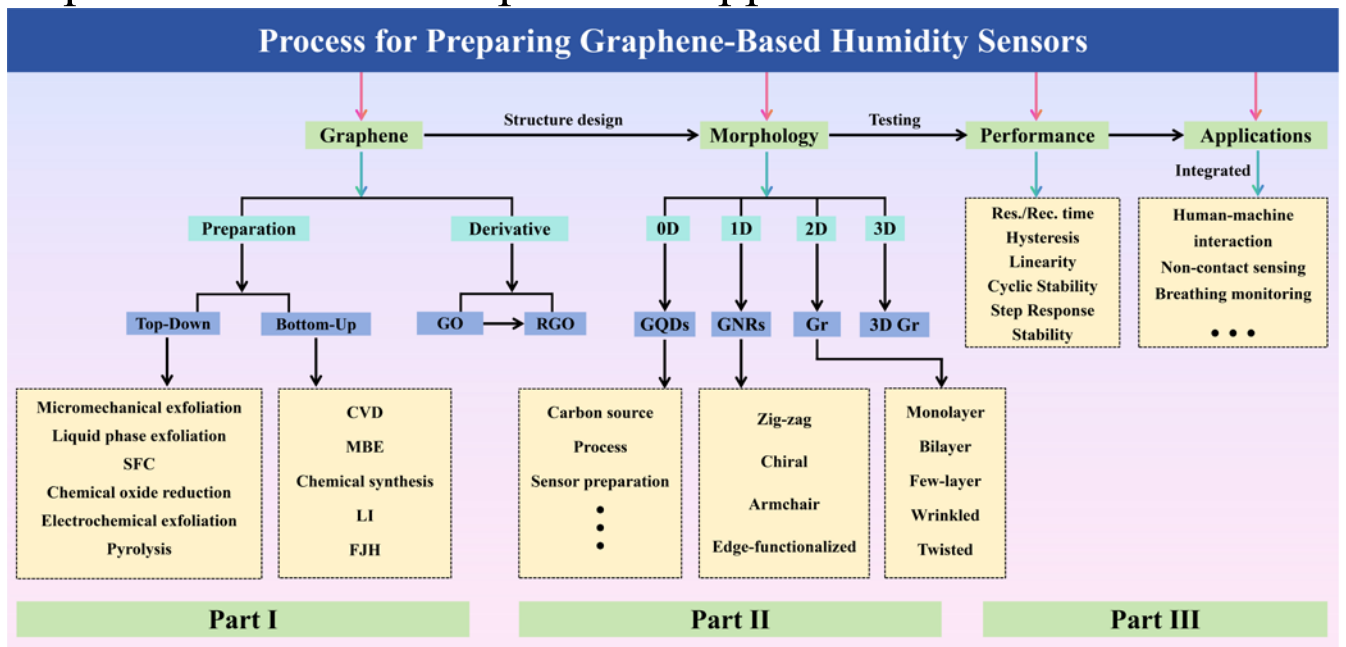

Fig. 1. Schematic diagram of preparation, structure design process and application process of graphene-based humidity sensor.

## II. Preparation of Graphene

Many innovative advances have been made in graphene over the past decades, and a variety of different graphene preparation techniques have been successfully developed, advancing the use of graphene materials in practical production. Two main methods are available for the preparation of graphene, one is to physically and chemically exfoliate the pristine graphite to fabricate monolayer or few-layer of graphene, also known as the "Top-down" method, and the other is to prepare graphene under high temperature and ultra-vacuum environment with the help of hydrocarbons, monocrystalline silicon and other materials containing element C as raw materials, which is also known as the "Bottom-up" method. In this section, we systematically review the two main types of graphene preparation along with their development history, and analyze their advantages and disadvantages of the two methods, respectively.

### *A. Top-Down*

The "Top-down" methods to prepare graphene mainly includes: Micromechanical exfoliation method, liquid phase exfoliation, supercritical fluid exfoliation (SFC) method, chemical oxide reduction (Hummers' method), electrochemical exfoliation method and pyrolysis.

In these methods, micromechanical exfoliation, liquid phase exfoliation and SFC method utilize mechanical external forces to overcome the van der Waals forces between graphite to exfoliate graphite layered crystals to prepare monolayer or few-layer graphene. In 2004, Geim and Novoselov's group [96] successfully exfoliated and observed monolayer graphene from highly oriented pyrolytic graphite (HOPG) for the first time by micromechanical exfoliation. Using this method, Geim's group successfully prepared quasi two-dimensional graphene and observed its morphology, revealing the reason for the existence of graphene's two-dimensional crystal structure (Fig. 2a). Liquid phase exfoliation and micromechanical exfoliation are commonly used physical methods for the preparation of graphene, but liquid phase exfoliation can handle relatively large amounts of graphite. A dilemma exists, however, in that suitable liquid phase systems (e.g., organic solvent, surfactant, etc.) and sufficient energy (e.g., ultrasound, high-speed shear, etc.) are needed to disrupt the van der Waals forces between the graphite. In 2008, Coleman's group [97] dispersed graphite in N-methyl pyrrolidone (NMP) for ultrasonication, successfully achieving liquid-phase exfoliation of graphite for the first time. The concentration of graphene obtained after centrifugation was 0.01 mg/mL, and the yield of monolayer graphene was 1%. The characterization results demonstrated that the obtained graphene was nearly free of defects. Furthermore, the group vividly proved that the transition from liquid-phase exfoliated graphite flakes to graphene occurs in three different stages in 2020 [98] (Fig. 2b).

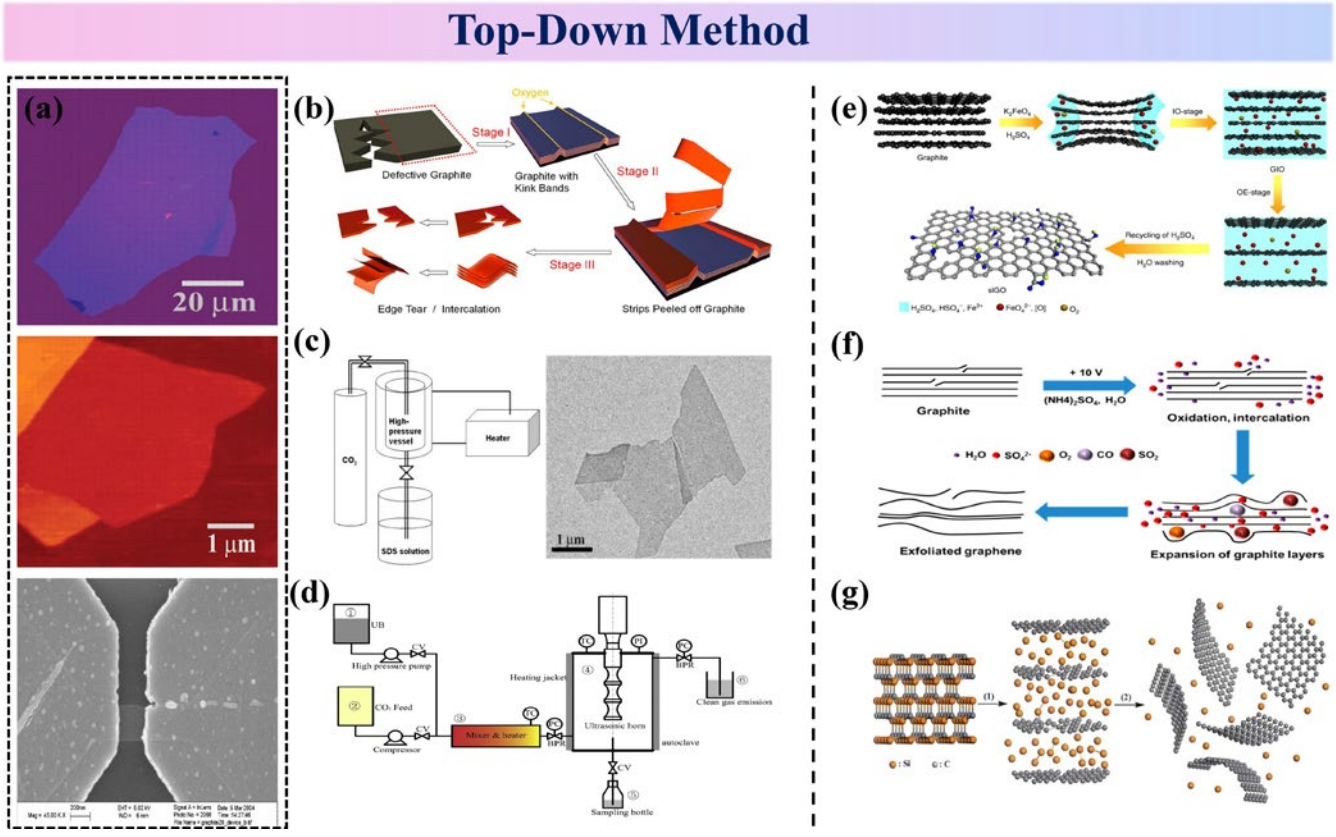

Fig. 2. Top-down preparation method of graphene. (a) Morphology of graphene films. (b) The process of liquid phase exfoliation of graphene. (c) Supercritical $CO_2$ exfoliation for the preparation of few-layer graphene process and graphene morphology. (d) Graphene preparation process by supercritical fluid (SCF). (e) Schematic diagram of graphene prepared by strong oxidant ($K_2FeO_4$). (f) Schematic diagram of electrochemical exfoliation of graphene. (g) Schematic diagram of monolayer graphene prepared from polycrystalline SiC particles. Adapted from [93], [95], [96], [100], [119], [121] and [122] under a Creative Commons license.

Supercritical fluid (SCF), combining the excellent properties of gases and liquids, features low interfacial tension, superior surface wettability as well as high diffusivity, which can be used as an intercalation agent to insert into the interlayer gap of graphite and exfoliate it into graphene. Pu et al. [99] firstly reported the preparation of graphene using supercritical carbon dioxide intercalation (Fig. 2c). Graphene was prepared by immersing natural graphite in supercritical carbon dioxide, and subsequently expanding and exfoliating the graphite through rapid decompression expansion. Atomic force microscopy results showed that the typical graphene sheets as-prepared contained about 10 atomic layers. Since then, the SCF method has been gradually promoted, and numerous new routes (e.g., fluid shear [100, 101], ultrasonic method [102-104] (Fig. 2d), ball milling [105, 106], organic molecule-assisted method [107-109], etc.) for preparing graphene based on the SCF method have emerged, leading to a prominent improvement in graphene yield and quality.

These methods above, however, exhibit typical drawbacks, such as poor controllability, small graphene size, low efficiency, high cost, and limited large-scale production. For further reducing the production cost of graphene and improving the production efficiency and utilization rate of graphene, chemical oxidation reduction method has been widely used for the preparation of graphene. The principle of chemical oxidation is the oxidation of graphite in strong acidic medium, leading to the destruction of its conjugate structure [110, 111], and introducing oxygen-containing functional groups (e.g., hydroxyl, carboxyl, etc.) between the graphite layers to obtain graphene oxide (GO). GO is finally obtained by removing oxygen-containing functional groups through thermal reduction [112-114], chemical reduction [115, 116], or electro-reduction [117] methods. Presently, three main oxidation methods have been developed, including Brodie method [118], Standenmaier method [119] and Hummers method [120]. Since the first two methods contain toxic gases (e.g., nitrogen, chlorine, etc.) in their preparation, the Hummers method has become the most commonly used method for graphite oxidation. However, the problems such as waste liquid pollution and structural defects during the preparation of graphene by chemical oxidation reduction method severely limits the large-scale preparation of graphene. Many efforts have been paid to solve above problems, for example, in 2010, Tour et al. [121] reported a method to oxidize graphite in mixed acid (sulfuric/phosphoric acid) medium with potassium permanganate as oxidant. This method has a higher oxidation efficiency and a more regular GO structure, which effectively solves the problem of structural defects in GO. The oxidizing agent is the most important control factor in the GO preparation process, however, the by-products of conventional oxidizing agents (e.g., $KClO_3$ and $KMnO_4$) are highly polluting and the intermediates pose an explosion risk. In order to develop a new oxidizing agent with high oxidation efficiency, no explosion risk, non-toxicity and non-pollution, Gao et al. [122] proposed an environmentally friendly and highly efficient green oxidizing agent, $K_2FeO_4$. Highly water-soluble GO can be obtained by using concentrated sulfuric acid for stirring, reacting at room temperature for 1 h, and then stirring the graphite flakes in the reactor, washing, and purifying. This strategy not only avoids the introduction of polluting heavy metals and toxic gases in preparation and products, but also recycles sulfuric acid, effectively solving the pollution problem (Fig. 2e).

In recent years, the technology of preparing graphene by electrochemical exfoliation has developed rapidly. This method provides the advantages of fast reaction, low pollution and low cost, besides, different thicknesses of graphene nanosheets can be obtained by controlling the composition of the electrolyte. In general, the electrochemical exfoliation method can be divided into anodic, cathodic and bipolar exfoliation [123]. The conventional electrochemical exfoliation method mainly takes graphite as the anode, different acid, alkali and salt solutions as the electrolyte system. Ions or charged molecules migrate into the graphite layer space with the help of electric current to achieve the purpose of stripping. For a specific example, Müllen et al. [124] used graphite as anode and platinum as cathode for electrochemical exfoliation in aqueous solutions of different inorganic salts ($(NH_4)_2SO_4$, $Na_2SO_4$, $K_2SO_4$, etc.). The results demonstrate that the as-prepared graphene (≤3 layers) shows high yields (>85%) and large lateral sizes (up to 44 μm) (Fig. 2f).

To ensure the production of highly pure, scalable, large-area graphene sheets, pyrolysis of silicon carbide is quite feasible. However, graphene obtained by epitaxial growth methods is located on the surface of single crystals and the yield of graphene is rather low. Therefore, the synthesis of freestanding monolayers of graphene in large quantities in the non-liquid phase remains a challenge. In this regard, Bao et al. [125] developed an efficient method to prepare graphene from commercially available polycrystalline SiC particles. The synthesis method involves heating commercial polycrystalline SiC particles in a tantalum reactor in a vacuum furnace. At 850 °C, Si began to be detected by mass spectrometry (MS), indicating that SiC began to decompose. After Si-C bond breaking and Si sublimation, the C elements self-assembled into individual monolayer graphene nanosheets at high temperature (Fig. 2g). STM results showed that the prepared graphene sheets exhibited a size of ~50-300 nm and a step height of 3.0 ± 0.4 Å, which closely matched the thickness of the monolayer graphene, featuring fewer defects compared to the graphene prepared by liquid-phase exfoliation. In general, the "Top-down" methods mainly involve the destruction of van der Waals forces between graphite layers via mechanical and chemical effects to obtain graphene. Each preparation method presents their own advantages and disadvantages, researchers should make a reasonable choice according to the experimental demands as well as conditions. TABLE II presents the analysis of advantages and disadvantages of different graphene preparation methods.

## B. Bottom-Up

The "Bottom-up" methods provide a means of constructing graphene through control at the atomic or molecular level. The "Bottom-up" methods typically focus on precisely controlling the formation of graphene to obtain specific structures and properties, mainly including chemical vapor deposition (CVD), molecular beam epitaxy (MBE), chemical synthesis, laser induction (LI) and Flash Joule Heating (FJH).

The main process of CVD method involves three processes: dissolution of carbon atoms onto a metal substrate, cooling and precipitation, and formation of graphene layer. In the 1970s [126], researchers used the CVD method to prepare graphene on a single-crystal Ni substrate, however, the quality and yield of the graphene was not guaranteed. In 2009, J. Kong and B.H. Hong et al. [127, 128] used Si wafers with polycrystalline Ni films as the substrate to prepare large area and few layers of graphene that could be completely transferred from the substrate, sparking a wave of graphene preparation by CVD method (Fig. 3a). Due to the problems of small grain size and defects in graphene grown using Ni film, in 2009, R.S. Ruoff et al. [129] used Cu substrate to grow monolayer graphene in a large-scale (Fig. 3b). According to the solubility difference of carbon atoms on different substrates [130, 131], the content of the prepared monolayer graphene was over 95%. Since then, high-purity Cu foil has been used as a mainstream substrate for the preparation of graphene through CVD method. However, the grain boundaries generally exist in the CVD-grown graphene and other 2D materials, which would impact the mechanical, electrical and chemical properties of CVD-grown graphene and other 2D materials [132, 133].

Molecular beam epitaxy (MBE) is a method to grow another crystal by lattice matching on one crystal structure, mainly including metal epitaxial growth, SiC epitaxial growth [134] and orientation attachment [135]. As early as in the 1970s, A.J. Van Bommel et al. [136] found that the surface of SiC was easily covered by a "graphite layer" through silicon evaporation during heat treatment of SiC, and the graphite layer had an obvious crystallographic relationship with the SiC crystals. A.J. Van Bommel speculated on the initial graphitization mechanism of SiC, opening the door for the preparation of graphene by MBE. Compared to chemical vapor deposition, MBE yields higher-quality graphene and allows for atomic-level control of graphene growth. However, the preparation methods of graphene by CVD and MBE require stringent conditions, such as suitable carbon sources (methane [137], ethylene, etc.), proper substrates (SiC, nickel, copper, cobalt, iron, etc.), and favorable growth conditions (pressure, temperature), which significantly increase the production cost and transfer efficiency of graphene.

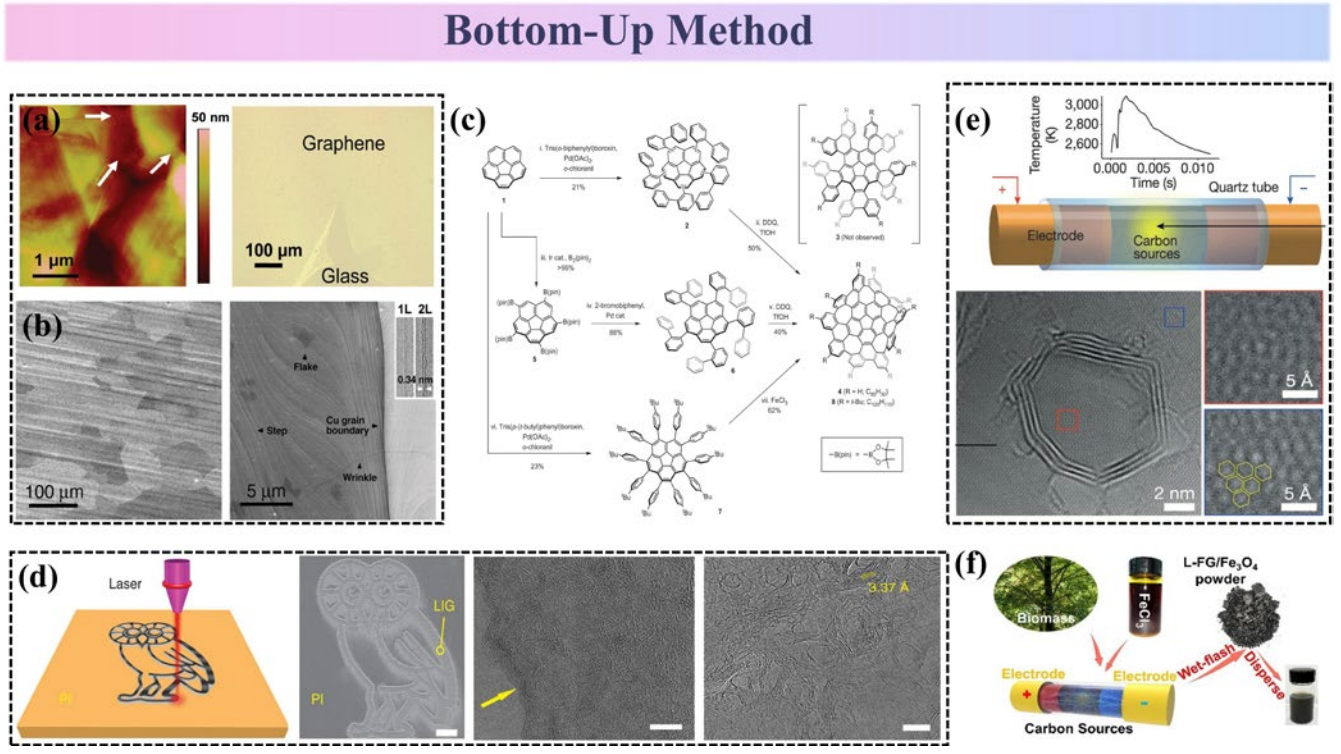

Fig. 3. Bottom-up preparation method of graphene. (a) Morphology of graphene prepared by CVD method. (b) Cu substrate to grow monolayer large-scale graphene. (c) Synthesis route to highly distorted graphene nanosheets. (d) Preparation of 3D porous graphene by laser-induced method. (e) Preparation of gram-scale graphene by FJH. (f) Schematic diagram of graphene preparation by FJH from lignin. Adapted from [124], [126], [136], [137], [147] and [151] under a Creative Commons license.

Another well-established method is chemical synthesis, which is a bottom-up assembly method for graphene synthesis. The chemical synthesis method starts from a small molecule with a precise structure and subsequently undergoes a controlled chemical reaction to obtain graphene with a well-defined structure [138]. Presently, graphene ribbons, graphene nanosheets, macroscopic graphene and their derived carbon-rich materials have been synthesized with polycyclic aromatic hydrocarbons (PAH) as precursors. For instance, Scott et al. [139] synthesized a highly distorted graphene nanosheet containing a five-membered ring in the center and surrounded by five seven-membered rings (Fig. 3c). $C_{80}H_{30}$ nanographene was synthesized by a stepwise chemical approach, which started from commercially available corannulene and was rapidly synthesized in two steps by C-H activation reaction. Experimental data show that the properties of this large graphene subgroup are affected by multiple odd ring defects. The highly distorted graphene nanosheets show wide bandgap, high fluorescence properties and are susceptible to redox reactions, further altering their electronic and optical properties. The advantages of this method include structural manipulation of graphene at the molecular scale and high processability, yet the resulting graphene exhibits smaller lateral dimensions and lower yields.

Significant work has also been devoted to optimizing the physical property changes that occur during the successful preparation of graphene. A relatively new, though truly advancing development was the work of Tour 's group in 2014 [140] showing that laser induction (LI) can be used for the preparation of graphene (LIG). Specifically, Tour's group found that 3D porous graphene can be generated by laser direct writing on the surface of a commercial polyimide (PI) film using $CO_2$ infrared laser system (Fig. 3d). The shape of the LIG can be drawn through the computer, providing a simple strategy for printable electronic devices. Moreover, The LIG, with the regulation of precursor [141-143], light source [144-146], and laser power [147, 148], can realize the controllable regulation of graphene from macro-scale to micro-scale. Despite the advantages of the LIG method, such as easy process, low cost, and controllable pattern shape, the adhesion ability of flexible precursors (e.g., polyimide films) is still not strong, and the appropriate choice of precursors significantly hinders the preparation efficiency.

In 2018, Hu et al. [149] utilized the carbothermal shock method to synthesize high-entropy-alloy nanoparticles (HEA-NPs) through a two-step rapid temperature rise and fall phase. Specifically, high voltage pulses can be used to bring the precursor-loaded carbon nanofibers to a high temperature environment of about 2,000 K in a very short period of time, with a rate of temperature rise and fall of ~105 K $s^{-1}$. Inspired by the work of LIG and Hu's group, in 2020, Tour's group [150] utilized this strategy to rapidly heat carbon black, which, to their surprise, turned into high-quality graphene (FG). Specifically, the carbon black was placed in a quartz or ceramic sample tube using a capacitor bank to apply high voltage, and the carbon black achieved temperatures above 3000 K in 100 milliseconds (Fig. 3e). This method of preparing one gram of graphene in just one second is known as Flash Joule Heating (FJH). The 2D peak is used to characterize the stacking of carbon atoms in a graphene sample, and the position of the G peak can determine the number of layers in a particular graphene sample. The higher the ratio of 2D/G peaks, the lower the number of layers of graphene. Surprisingly, graphene prepared by FJH method has a ratio of $I_{2D}$/G as high as 17 (The 2D and G peaks are used to determine the number of graphene layers. The higher the ratio of 2D/G, the fewer layers of graphene), exhibiting the highest value of graphene prepared by any method reported so far. Furthermore, the FJH method offers the advantages of abundant carbon sources (e.g., waste plastics [151, 152], rubber tires [153], lignin [154] (Fig. 3f), etc.), low cost (electrical energy consumption of ~7.2 kJ g-1), low pollution (solvent-free method), and high yield (gram-scale graphene produced in one second), making it suitable for various application requirements.

TABLE I
COMPARISON OF ADVANTAGES AND DISADVANTAGES OF DIFFERENT GRAPHENE PREPARATION METHODS.

| Preparation Methods of graphene | Type | Advantages | Disadvantages | Ref. |
|---|---|---|---|---|
| Micro mechanical exfoliation | Top-Down | Simple, high quality | Poor controllability, small size, low efficiency, high cost, limited large-scale production | [96] |
| Liquid phase exfoliation | Top-Down | Simple process, high quality, low defects | Low yield, low rate for graphene monolayer, small size, environmental pollution, long preparation period | [97] |
| SFC | Top-Down | Simple, low cost | High-pressure equipment is required, low yield of monolayer graphene | [99] |
| Chemical oxide reduction | Top-Down | Low cost, high productivity, large-scale production | Defects and impurities, small size, environmental pollution | [121] |
| Electrochemical exfoliation | Top-Down | Low cost, short reaction time, mild reaction conditions, green chemistry, controllable number of layers | Defects and impurities, limited large-scale production | [124] |
| Pyrolysis | Bottom-Up | High quality, high controllability, wide applicability, fast growth rate | High temperature environment, high cost, complex equipment, lattice defects | [125] |
| CVD | Bottom-Up | High quality, controllable number of graphene layers, scalability, preparation of large-area graphene films | Complex process, high cost, grain boundaries, in need of transfer | [129] |
| MBE | Bottom-Up | Large area, high quality and high efficiency | Uncontrollable number of graphene layers, expensive raw materials, lattice matching, immature technology | [136] |
| Chemical synthesis | Bottom-Up | Versatility, functionalization, graphene molecular scale manipulability, processability, low cost | Long reaction time, environmental pollution, complex process | [139] |
| LI | Bottom-Up | High resolution, wide applicability, controllability, fast preparation | Limited to small size structures, slow preparation speed, high equipment cost | [140] |
| FJH | Bottom-Up | Fast preparation, high purity, controllability, high solubility, low defects | High energy consumption, limited to precursor materials, limited large-scale preparation | [154] |

## III. CONFIGURATION AND MECHANISM OF HUMIDITY SENSORS

Different advanced fabrication methods lead to different sensor structures and mechanisms. Typical humidity sensor fabrication techniques involve ceramic-based fabrication (using metal oxides as raw materials to make porous ceramics through the ceramic process), semiconductor fabrication (utilizing the properties of silicon or other semiconductor materials to form sensitive components with the help of micro-nano-processing techniques, such as mechanical stripping, photolithography, etching, ion implantation, etc.), film technology (spin-coating, spray-coating, drop-coating, casting, vacuum filtration, inkjet printing, etc.), and other advanced fabrication techniques (freeze-drying, polymer cross-linking, 3D/4D printing, laser direct writing, electrospinning, etc.) Depending on the manufacturing technology, humidity sensors with different operating mechanisms can be obtained, such as resistive, capacitive, impedance, SAW, QCM, FET and fiber optic humidity sensors.

### A. Resistive Humidity Sensors

Graphene resistive humidity sensor has been one of the most widely studied humidity sensing devices due to its advantages of simple fabrication, low cost and stability. The abundant oxygen-containing functional groups on the surface of graphene can adsorb water molecules through hydrogen bonding and van der Waals forces. The adsorbed water molecules affect the electronic structure of graphene and change its electrical conductivity. Specifically, water molecules adsorbed on the surface of graphene affect the electronic density of states and carrier concentration of graphene, thereby changing the resistivity of graphene. The sensitivity of a resistive humidity sensor can be defined as [155]:

$$\text{Response} = R\text{-}R_o/R_o \times 100\% \quad (1)$$

where R is the resistance measured under experimental conditions and $R_0$ is the resistance measured under dry air conditions. Presently, the commonly accepted mechanism for graphene resistive humidity sensors is the Grotthuss chain reaction [155] ($H_2O + H^+ \rightarrow H_3O^+$ and $H_2O + H_3O^+ \rightarrow H_3O^+ + H_2O$).

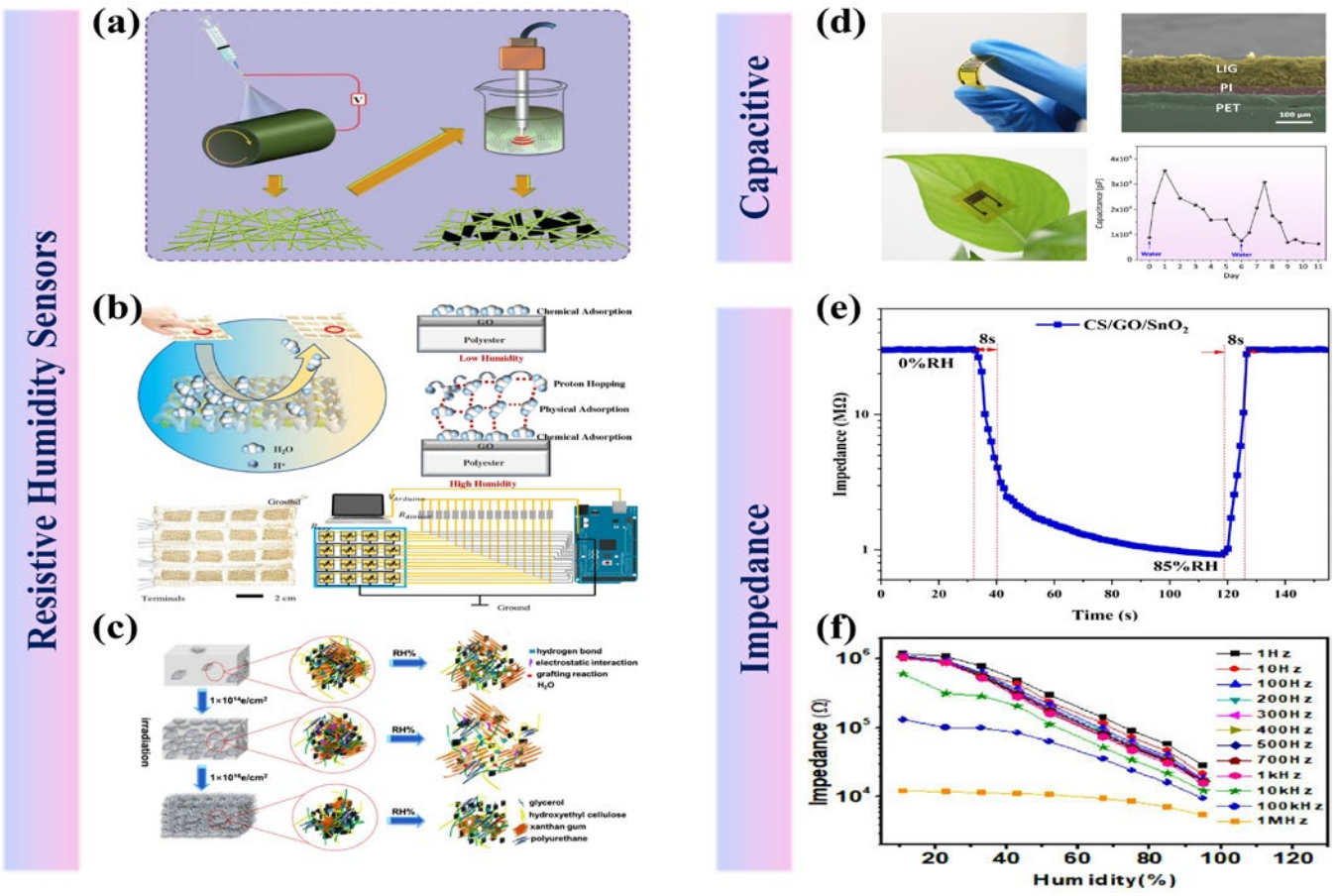

Fig. 4. Different sensing mechanisms of humidity sensor. (a) Schematic diagram of graphene resistive humidity sensor prepared by electrostatic spinning. (b) Functionalized humidity sensors prepared on polyester fibers using the coating method. (c) Preparation of graphene/polymer humidity sensors with low crosslink density by electron irradiation method. (d) GO flexible moisture-sensitive sensors for detecting plant transpiration based on capacitive-sensing. (e) Response/recovery time of GO/CS/$SnO_2$ impedance humidity sensor. (f) Impedance-frequency test plot of the rGO-$BiVO_4$ heterojunction humidity sensor. Adapted from [153], [154], [155], [156], [157] and [158] under a Creative Commons license.

Graphene resistance humidity sensors usually integrate humidity sensitive materials into electrodes to form conductive channels. Many advanced methods are proposed to prepare graphene humidity sensors, for instance, Yang et al. [156] applied ultrasound to embed graphene into a nylon 66 film prepared by electrostatic spinning to design a high-performance humidity sensor with a fast response/recovery time (35/90 s) and a change in conductance of more than 30% when increasing relative humidity (RH) from 0 to 90% (Fig. 4a). Flexible fiber electronics have a wide range of applications in wearable devices, healthcare, and energy harvesting and storage. For example, Chen et al. [157] prepared functionalized humidity sensors by coating GO onto polyester fibers for an e-textile keyboard (Fig. 4b). Since the humidity field value near human fingertips is much higher than other parts of human body, thus, the signal can be detected in a unimodal and non-contact manner by this e-textile keyboard. Irradiation is a simple, fast and green processing technique to functionalize materials by producing various defects and free radicals. Recently, Li et al. [158] fabricated graphene/polymer composites with low cross-linking density through electron irradiation. Among them, graphene and polyimide films were used as conductive materials and flexible substrates, respectively. A mixture of cellulose, xanthan gum, glycerol and polyurethane is used as a sensitive material. The results indicated that the sensitive materials are prone to form unique network structures, folds, defects, and hydrophilic functional groups at suitable irradiation energy density (Fig. 4c), which significantly enhance the moisture absorption ability of the sensitive materials.

### B. *Capacitive Humidity Sensors*

The working principle of capacitive humidity sensors mainly depends on the influence of water molecules on the dielectric constant. As the humidity changes, the content of water molecules in the air changes, thus affecting the capacitance value of the sensor. The response of the sensor can be defined by the following formula:

$$\text{Response} = C\text{-}C_0/C_0 \times 100\% \quad (2)$$

where C and $C_0$ represent the capacitance measured under measured humidity and dry air, respectively.

In addition, the capacitance value can further be defined by the following formula:

$$C = \varepsilon A / d \quad (3)$$

where ε, A and d represent the dielectric constant, the area of parallel capacitance and the distance between parallel plates of the capacitor, respectively. Capacitive humidity sensors are mainly composed of two parts: a moisture-sensitive capacitor and a conversion circuit. The moisture-sensitive capacitor is composed of two parallel metal electrodes and a moisture-sensitive material sandwiched between them. The dielectric constant (ε) of this moisture-sensitive material changes with the ambient relative humidity. When the ambient humidity changes, the capacitance of the moisture-sensitive element also changes. Generally, when the relative humidity increases, the moisture-sensitive capacitance increases; when the relative humidity decreases, the moisture-sensitive capacitance decreases. In addition, the distance (d) and area (A) of the metal electrodes also affect the moisture-sensitive capacitance. The change in the humidity-sensitive capacitance is converted into a change in voltage through a conversion circuit, so that the change in relative humidity can be reflected by the change in voltage. The conversion circuit can be a simple RC circuit or a more complex analog or digital circuit for signal amplification, filtering and linearization. For example, Li et al. [159] first prepared graphene interdigital electrodes on PI substrates by laser direct writing, and then covered the electrodes with GO as a humidity sensitive material to prepare flexible humidity sensors. The prepared flexible humidity sensor exhibits low hysteresis, high sensitivity (3215.25 pF/%RH) and long-term stability which can be attached to the plant leaves to track the transpiration of the plant without incurring any damage to the plant (Fig. 4d).

### C. *Impedance Humidity Sensors*

Impedance type humidity sensors are another commonly used type of sensor. The impedance output of impedance type humidity sensors is highly dependent on RH and frequency. As the RH and frequency decreases, the impedance significantly increases, thereby, the sensor exhibits better performance at lower frequency. Structurally, similar to resistive humidity sensor, impedance type humidity sensor consists of three parts: substrate, interdigital electrodes and sensitive materials. In principle, the impedance-type humidity sensor is a device that measures its output current by applying a sinusoidal voltage to the electrodes. Where, the impedance value is calculated by the ratio of voltage to current. For example, Yang et al. [160] proposed a solution mixing method to prepare impedance-type humidity sensors. Specifically, chitosan (CS) and $SnO_2$ were added to GO solution to prepare composite moisture-sensitive materials. The humidity sensor prepared from this composite material exhibited high response (72683%), high sensitivity (402.5 kΩ/%RH), and fast response/recovery time (8 s/8 s) (Fig. 4e). Moreover, He et al. [161] reported a humidity sensor based on rGO-$BiVO_4$ heterojunction nanocomposites. The heterojunction formation was demonstrated by Mott-Schottky plot, and the narrow bandgap of the composite was demonstrated to be favorable for electron transfer by means of UV-vis absorption spectroscopy. Impedance-frequency tests indicated that with the increase of RH, the frequency characteristics of the sensor is almost unaffected from 1 Hz to 1 kHz (Fig. 4f).

## D. Surface Acoustic Wave (SAW) Humidity Sensors

A SAW humidity sensor is a device that utilizes the principle of sound waves propagating on surfaces to measure humidity. As the humidity changes, the moisture-sensitive material of the sensor absorbs or releases water molecules, leading to changes in the propagation speed of sound waves [162]. Structurally, the SAW humidity sensor mainly comprises surface acoustic wave wafer, moisture sensitive layer, input interdigitated transducer (IDT), signal processing circuit, and protective layer. Surface acoustic wave (SAW) humidity sensor based on graphene typically utilizes graphene material as the acoustic layer. The affinity of graphene material to water molecules changes in mass, leading to variations in SAW (mass loading effect). For example, Kim et al. [163] presented a humidity sensor based on multifrequency surface acoustic wave resonance (SAWR). GO was deposited into the SAWR region by electrostatic spray deposition (ESD) (Fig. 5a). The results demonstrated that the sensitivity and stability of the sensor were highly dependent on the resonance frequency. By adjusting the appropriate operating frequency, the stability and sensitivity of the sensor in terms of frequency shift and quality factor (Q) were improved by 150% and 75%, respectively.

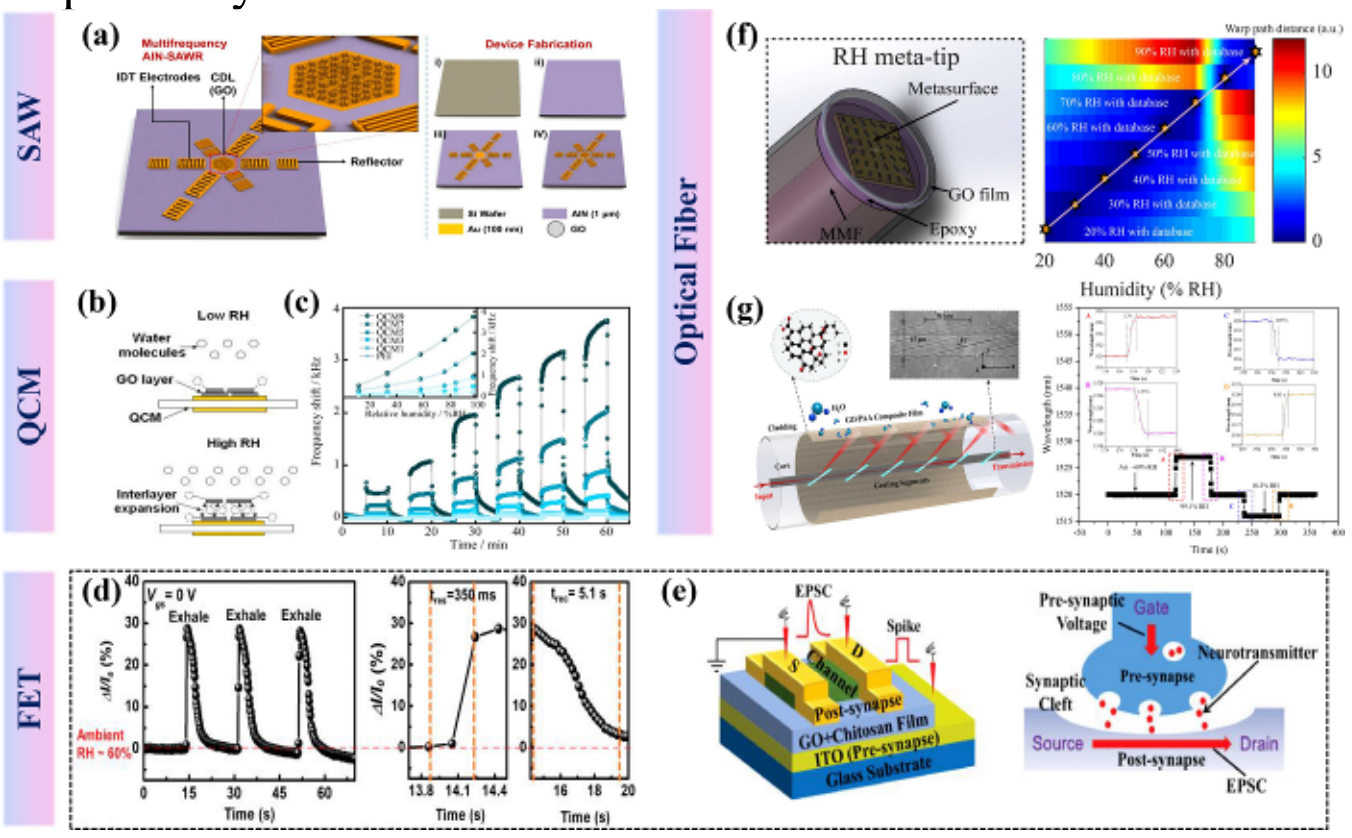

Fig. 5. Different types of mechanisms of humidity sensor. (a) Schematic of a multi-frequency surface acoustic wave resonance (SAWR) humidity sensor. (b) Humidity sensor prepared by coating GO on QCM. (c) Step response curves of humidity sensors prepared by depositing GO/PEI on QCM electrodes. (d) Respiration monitoring graph of graphene field effect transistor (GFET) humidity sensors with response/recovery time. (e) Schematic diagram of IZO-based synaptic transistor for GO/CS humidity sensor. (f) Schematic of a fiber-optic humidity sensor combining a moisture-sensitive meta-tip with optical barcoding technology. (g) Schematic and response/recovery time curve of the GO/PAA fiber optic humidity sensor. Adapted from [160], [162], [163], [164], [165] , [166] and [167] under a Creative Commons license.

TABLE II
SUMMARY OF SENSING PERFORMANCE AND PREPARATION METHODS OF GRAPHENE-BASED HUMIDITY SENSORS.

| Sensor type | Sensing materials | Preparation of sensors | Sensitivity/Response | Res./Rec. time | Meas. range | Year | Ref. |
|---|---|---|---|---|---|---|---|
| | Borophene/graphene | Drop coating | 4200% | 10.5/8.3 s | 0-85% | 2021 | [164] |
| | GQDs | Drop coating | - | 15/55 s | 10-90% | 2020 | [165] |
| | PVA/graphene | Electrospinning | 66.4% | 11/35 s | 10-80% | 2020 | [166] |
| | Graphene carbon ink /paper | Screen printing | ~12.4 Ω/%RH | ~4/~6 s | 25-91.7% | 2022 | [167] |
| Resistance | Coolmax/Graphene-Oxide | Spinning and weaving | ~83% | 0.5/0.6 s | 45-90% | 2021 | [168] |
| | rGO/$WS_2$ | Spray coating | ~17% | 31/95 s | 0-91.5% | 2021 | [169] |
| | rGO/PANI | Layer by layer | 37.4% | 61/62 s | 15-95% | 2023 | [170] |
| | rGO | Spray coating | 0.33%/RH | 25/127 ms | 0-100 s | 2020 | [171] |
| | SF/rGO | Screen printing | 60% | 21.51/85.62 s | 6-90% | 2023 | [172] |
| | rGO/$MoS_2$ | Drop casting | 0.973%/%RH | 4/15 s | 30-80% | 2021 | [40] |
| | Cellulose fiber/GO | Layer by layer | 1442500% | 1.3/0.8 s | 10-90% | 2023 | [173] |
| Capacitance | GOQDs /CNF | Vacuum freeze drying | 51840.91 pF/%RH | 30/11 s | 11–97% | 2022 | [174] |
| | GO | Drop casting | 3215 pF/%RH | 15.8/- s | 20-80% | 2020 | [159] |
| | LISG/GO | Laser induced | - | 58/8 s | 11-97% | 2022 | [175] |
| | GOQDs | Drop casting | 1025.99 pF/%RH | 2.2/1 s | 22-97.3% | 2021 | [176] |
| | Graphene printing ink | Inkjet printing | 0.03 pF/%RH | 2.46/2.63 s | 40-70% | 2023 | [177] |
| | Porous Anodic Alumina/ GO | Drop casting | 1178.76 pF/%RH | 0.8/16 s | 0.9-99% | 2023 | [178] |
| | $GO/MoTe_2$ | Drop casting | 94.12 pF/%RH | 39/12 s | 11.3–97.3% | 2023 | [179] |
| | $PPy/BiPO_4$ | Dip coating | 24.6% | 340/60 s | 12-90% | 2021 | [180] |
| | CMC/graphene | Drop casting | 97% | ~300/- s | 11-95% | 2023 | [181] |
| | LIG/ PEEK | Ion sputtering | 1700 KΩ/%RH /200 Hz | 4.2/6.8 s | 11-97% | 2022 | [182] |
| Impedance | $CS/GO/SnO_2$ | Drop coating | 402.5 kΩ/%RH | 8/8 s | 0-85% | 2023 | [160] |
| | PDDA/c-PVA/GO | Laser direct writing and In-situ crosslinking | - | 3/- s | 10-98% | 2023 | [183] |
| | NWs/GQDs | Dip coating | 40.16 kHz/%RH | 27/12 s | 20-90% | 2020 | [184] |
| | $Graphene/PVA/SiO_2$ | Spin Coating | −2.429 kHz/%RH | 24/14.4 s | 5-90% | 2020 | [184] |
| | $GQDs/PEI/SiO_2$ NPs | Drop coating | 2400 kHz/%RH | 20/5 s | 20-80% | 2023 | [185] |
| SAW | GO | Drop coating | 3.33 kHz/%RH | 80/72 s | 0-97.3% | 2020 | [186] |
| | $MoS_2/GO$ | Drop casting | 766.8 ppm/%RH | 6.6/3.5 s | 20-95% | 2022 | [187] |
| | GO | Vacuum filtration | 51 kHz/%RH | 9/12 s | 10-90% | 2020 | [188] |
| | PDA@CNC/GO | Drop casting | 54.66 Hz/%RH | 11/4 s | 11.3-97.3% | 2020 | [189] |
| | GO | Scrape coating | - | 10/11 s | 23-92% | 2020 | [190] |
| QCM | GO/carbon nanocoil | Drop casting | 4618 Hz/%RH | 4/2 s | 0-97% | 2020 | [191] |
| | GO/PEI | Dip coating | 37.84 Hz/%RH | 5/8 s | 11.3-97.3% | 2020 | [192] |
| | $MoS_2/GO/C_{60}$ | Drop coating | 31.8 Hz/%RH | 1.3/1.2 s | 2-97% | 2021 | [193] |
| | GQDs | lnk-jet printing | 78% | - | 3.4-81.3% | 2021 | [194] |
| FET | PS/rGO | Drop casting | - | ~60/~60 s | 45-90% | 2023 | [195] |
| | Graphene/ion gel | Drop casting | 45% | 0.35/5.1 s | 22-92% | 2021 | [196] |
| | Graphene | Wet etching transfer | 350 Hz/%RH | - | 0-100% | 2022 | [197] |
| | LIG | Dip coating, LI | 0.187 dB/%RH | 0.178/0.713 s | 5-95% | 2022 | [198] |
| Optical fiber | Polyelectrolyte/GO | Dip coating | 138.7 pm/%RH | 1.3/1.95 s | 8.3-99.1% | 2022 | [199] |
| | GO | Dip coating | 18.5 pm/%RH | 0.042/0.115 s | 30-80% | 2020 | [200] |

### *E. Quartz Crystal Microbalance (QCM) Humidity Sensors*

QCM humidity sensor is a type of sensor commonly used to measure humidity in gases or liquids. The quartz crystal in a QCM sensor is cut into thin slices and placed in a circuit. When a voltage is applied to a quartz crystal, alternating compression and stretching vibrations occur at the electrodes. [201]. In addition, water molecules adsorbed on the surface will change the mass of the quartz crystal, resulting in a change in its resonant frequency. By measuring the change in resonant frequency, the amount of water adsorbed on the quartz crystal can be determined, enabling the calculation of the humidity level in the environment. Presently, the mass effect and the viscosity effect between water molecules and the sensing material are commonly recognized as the two interactions in QCM humidity sensors. For example, Yao et al. [202] prepared a humidity sensor by coating GO onto QCM for the first time (Fig. 5b). The results showed that the humidity sensor exhibited high sensitivity (22.1 Hz/% RH) and excellent linearity ($R^2$=0.9718) in the range of 6.4-93.5% RH. They also found that within different high and low humidity ranges, the frequency response is closely associated with the mass change caused by water molecule adsorption/desorption and the expansion effect of graphene. Recently, Tai et al. [192] developed a facile chemical layer-by-layer self-assembly (CLS) method to deposit GO/polyethyleneimine (PEI) multilayers on QCM electrodes to prepare humidity sensors. The results showed that the GO/PEI humidity sensor exhibited high sensitivity (37.84 Hz/%RH) (Fig. 5c), fast response/recovery (< 5 s/8 s), and stable step response performance.

## F. Field-Effect Transistor (FET) Humidity Sensors

Graphene FET humidity sensor is mainly composed of source, drain, gate and channel materials, where graphene is usually used as a channel material to bridge the source and drain. The main mechanism of the graphene FET humidity sensor involves the sensitive layer absorbing or releasing moisture, resulting in changes in the charge state of the sensitive layer, which affects the charge distribution between the gate and the sensitive layer. In FET humidity sensors, the low on/off ratio and zero bandgap of graphene significantly limit the performance of the sensor. Fortunately, graphene can be modified by functionalization or compounded with other materials to improve the sensitivity of humidity sensors. For example, Liu et al. [196] prepared a graphene field effect transistor (GFETs) using wet-transferred graphene as the channel material, ion gel as the gate dielectric and humidity-sensing layer. The polar water molecules adsorbed by the ion gel induced the migration of ions in the gate dielectric, which in turn regulated the current output in the GFET. The results showed that the ion-gel graphene humidity sensor exhibited an outstanding sensitivity regime (the normalized current change per percent RH) (0.0135) even under high humidity (>71% RH). In addition, the humidity sensor can monitor human respiration in real time with a response time of 350 ms (Fig. 5d). In order to achieve higher sensitivity, faster response/recovery time, and lower energy consumption, Wan et al. [203] used GO/chitosan composite electrolyte as the gate dielectric layer to prepare IZO-based synaptic transistors for humidity sensors (Fig. 5e). The results showed that the energy consumption of the humidity sensor based on the synaptic transistor at 75% RH was 2.2 nJ and the response time was 30 ms.

## G. Optical Fiber Humidity Sensors

In optical fiber humidity sensors, the moisture content adsorbed by the high refractive index coating in the optical fiber affects the transmission of light, resulting in a difference between the detected optical signal and the reference optical signal, which can be used to calculate the air humidity. The optical fiber humidity sensors exhibit many advantages, such as high accuracy, fast response, resistance to external electromagnetic interference, and long service life. The optical response of GO to near-infrared light as well as its good hydrophilicity can be used as good intermediate substance between environmental humidity and optical fibers. For example, Huang et al. [204] developed a fiber-optic humidity sensor combining a humidity-sensitive meta-tip with optical barcode technology, which can recognize the change of relative humidity directly by human eyes. The humidity sensitive meta-tip consists of GO film and multimode fiber (MMF) integrated with metasurface (Fig. 5f). The results showed that the warp path distance (WPD) sequence is closely related to two different expansion coefficients of graphene under different humidity conditions. In addition, the barcodes corresponding to the WPD sequences arranged in humidity order displayed good human eye identification characteristics. Another crucial characteristic of optical fiber humidity sensors is the durability and hydrophilicity of the humidity-sensitive layer, which requires good adhesion of the humidity-sensitive material to ensure stability during measurement. For instance, a fiber optic humidity sensor was prepared by coating hydrophilic GO/poly (acrylic acid) (PAA) on an excessively tilted fiber grating [199] (ex-TFG) (Fig. 5g). The results showed that PAA provided good adhesion and a highly hydrophilic network on the optical fiber. The fiber optic humidity sensor exhibited high sensitivity (138.7 pm/%RH), fast response/recovery time (1.3/1.95 s), and good durability in the RH range of 8.3-99.1% RH, posing great potentials in the detection of human respiration and disease diagnosis. As shown in TABLE II and Fig. 6, we summarize the performance comparison results of graphene and its derivatives humidity sensors and the statistical data graph of research papers on different types of graphene sensor mechanisms over five years.

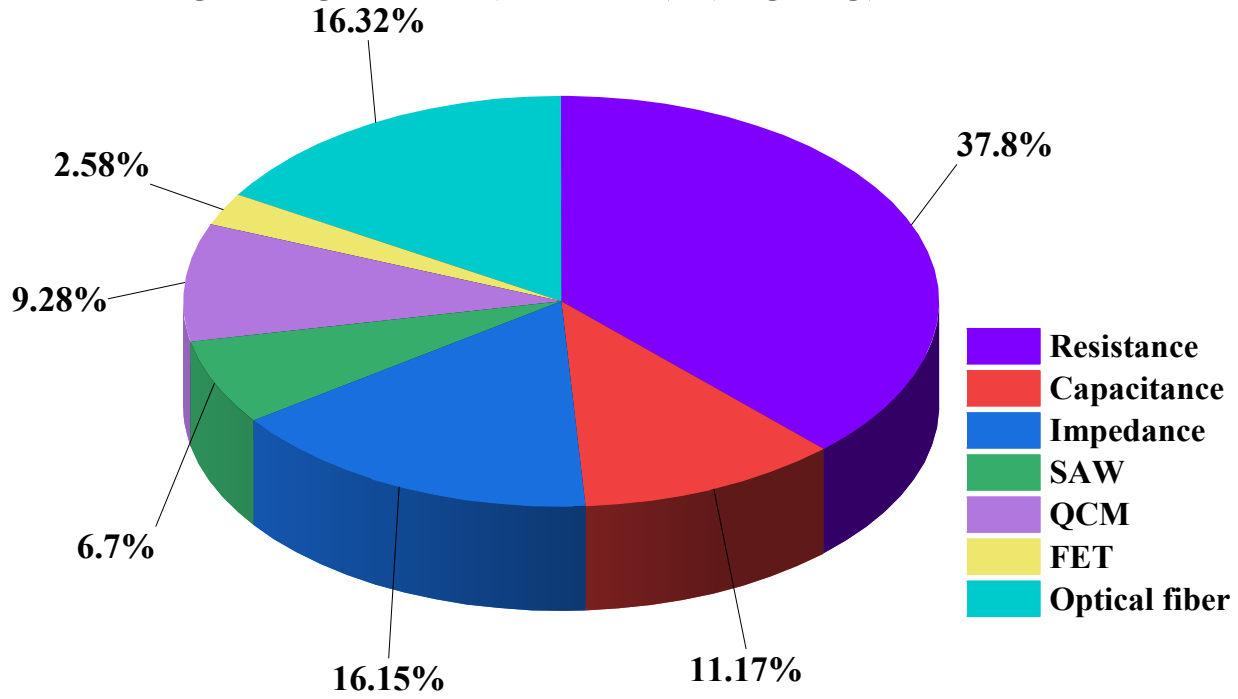

Fig. 6. The statistical pie chart of the number of research papers on the mechanism of different types of graphene humidity sensors from 2020 to 2024.

In addition, combining different operating principles or sensing mechanisms is also a way to improve the overall performance of the sensor. For example, common composite mechanisms for humidity sensors include:

(1) resistance-capacitance composite mechanism [205-207] (e.g., when humidity increases, water molecules may adsorb on the sensor surface, resulting in a change in the resistivity of the material, and at the same time the adsorption of the water molecules may change the capacitive properties of the material surface).

(2) Ion Migration-Proton Conduction Compound Mechanism [208] (In some humidity sensors based on ion-conducting materials, changes in humidity can lead to changes in ion mobility, while also altering the proton conduction properties within the material.).

(3) Optical-Electrical Compound Mechanism [209] (Humidity-induced changes in the refractive index or changes in spectral absorption properties are utilized to detect humidity, while the electrical sensor monitors the changes in resistance or capacitance.).

(4) Thermal conductivity-resistance composite mechanism [210] (Changes in humidity affect the thermal conductivity of the material and also affect the resistance of the material.)).

(5) Mechanical-electrical composite mechanism [211] (in flexible or stretchable humidity sensors, changes in humidity may cause mechanical deformation of the material, such as expansion or contraction).

(6) Chemical-physical sensitive composite mechanism [212] (some humidity sensors may combine chemically sensitive materials (such as certain specific polymers or metal-organic framework materials) and physically sensitive elements (such as nanomaterials or semiconductor materials)).

## IV. Structural Design of Graphene-Based Humidity Sensors

Graphene, as a unique carbon material, has been designed with different dimensional structures providing diversity in various fields. The properties of graphene, such as electrical conductivity, optical transmittance, thermal conductivity and mechanical strength, can be precisely controlled by structural design, allowing graphene to be tailored for specific applications and different fields. In this section, the structural design of graphene on zero-dimensional, one-dimensional, two-dimensional as well as three-dimensional scales will be discussed and its advantages in the field of humidity sensing will be thoroughly analyzed.

### *A. Zero-Dimensional Structure*

Zero-dimensional graphene structure is represented by nanoscale zero-dimensional dotted graphene, such as graphene quantum dots (GQDs). The advantage of zero-dimensional graphene structure design lies in the nanoscale size effect and quantum effect [213, 214], enabling electrons to exhibit unique properties in this dimension. GQDs exhibit a wide range of applications in biological[215], optoelectronic [216], environmental [217], and electronic fields, offering new possibilities for nano electronic devices. Resistive humidity sensors, as the most commonly used sensor type, exhibit the advantages of simple process and high sensitivity, yet large size and current drift issues severely limit their stability and multifunctional applications. To overcome this obstacle, one of the reliable methods is to rationally nanosize the moisture-sensitive materials. For instance, Lee et al. [194] developed a humidity sensor based on silicon metal-oxide-semiconductor field-effect transistors (MOSFETs). Graphene quantum dots, as moisture-sensitive materials, were locally deposited on MOSFET by inkjet printing, effectively reducing the consumption of GQDs solution (Fig. 7a). Furthermore, TEM and SEM results showed that the size of graphene quantum dots was around 2.01 nm, accompanied by significant quantum effects (Fig. 7b-c). Experimentally, the response and recovery characteristics of the sensor were improved by 40% and exhibited excellent long-term stability by applying a pre-bias of -1 V to the control-gate (Fig. 7d).

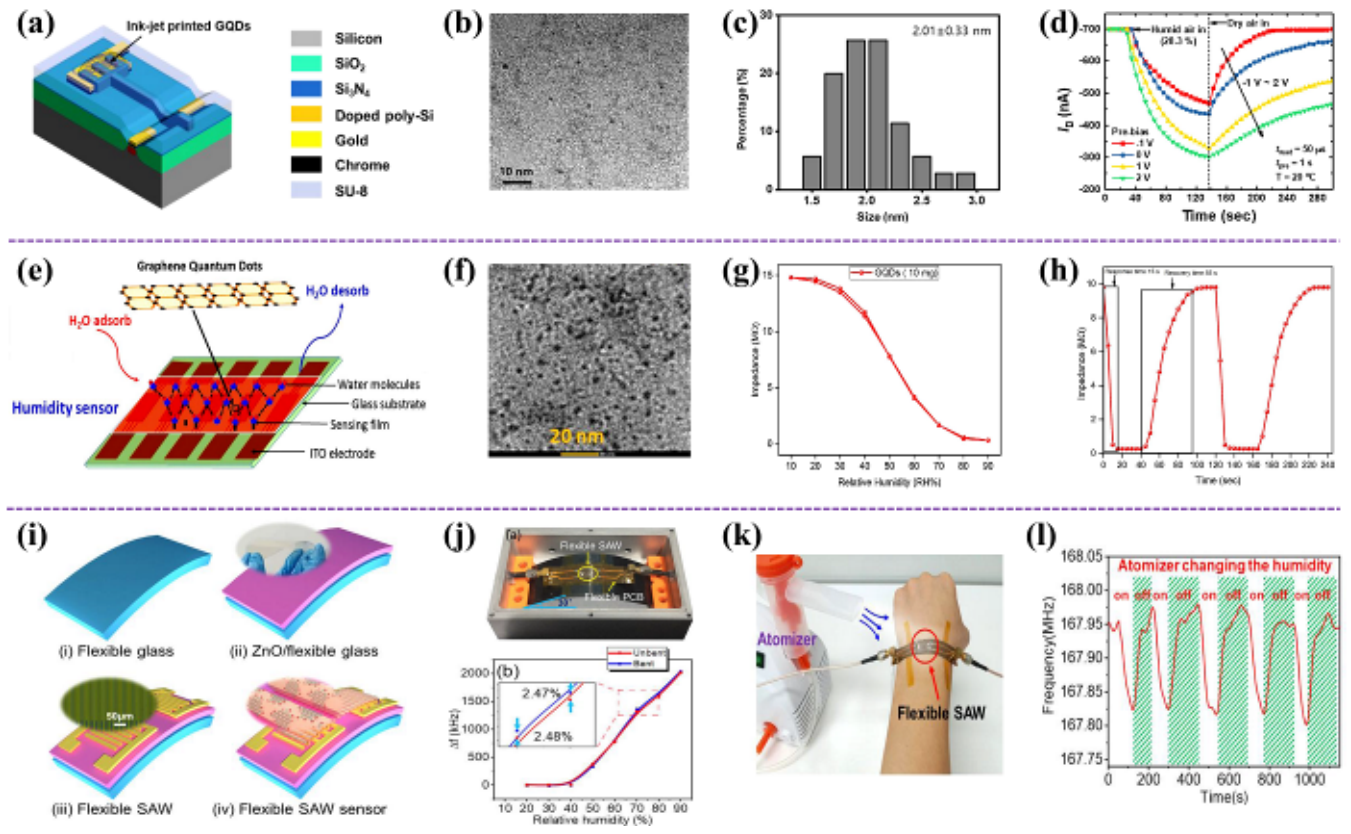

Fig. 7. Zero-dimensional structure of graphene-based humidity sensors. (a) Schematic of MOSFET-based graphene quantum dot humidity sensor. The (b) size, (c) morphology, and (d) response and recovery characteristics of graphene quantum dots. (e) Schematic diagram of GQDs humidity sensors prepared from graphite wastes. The (f) size, (g) hysteresis characteristics, and (h) response/recovery time of graphene quantum dots. (i) Transparent flexible SAW humidity sensor based on ZnO nanowires, GQDs and flexible glass. The (j) bending frequency drift characteristics, (k) human humidity detection, and (l) resonance frequency variation curves of SAW humidity sensors. Adapted from [198], [169] and [188] under a Creative Commons license.

Conventional methods for preparing GQDs include ultrasound-assisted [218], laser-assisted [219, 220], or chemical methods [221, 222] with expensive carbon sources such as carbon fibers [223], carbon nanotubes [224], and organic compounds [225]. Hence, an economical, efficient and facile method for preparing GQDs is urgently needed. A strategy for the preparation of GQDs humidity sensors from graphite waste was proposed [165], where GQDs were deposited on ITO/glass electrode substrates using spin-coating method (Fig. 7e). The TEM images demonstrated that the as-prepared GQDs exhibited regular shapes and sizes, with an average size of 10 ± 0.5 nm (Fig. 7f). Furthermore, the impedance humidity sensors based on graphene quantum dots showed a low hysteresis of 2.2%, and fast response times and recovery time (15 s and 55 s) (Fig. 7g-h). As reported above, the construction of graphene quantum dot humidity sensors cannot combine both flexibility and high sensing performance, significantly limiting the application of the sensors in wearable devices. Meanwhile, the poor adhesion of moisture-sensitive materials on flexible substrates for high-performance sensing is a great challenge [226]. Many efforts have been devoted to the preparation of flexible high-performance humidity sensors, However, they simply adhere to the surface of devices and neither form various spatial three-dimensional structures nor increase the number of active sites significantly [227]. Recently, Zhou et al. [184] developed a transparent flexible SAW humidity sensor based on ZnO nanowires (NWs), GQDs and flexible glass (Fig. 7i). The high sensitivity (40.16 kHz/%RH) and excellent stability of this humidity sensor are attributed to the large specific surface area of ZnO NWs and numerous hydrophilic functional groups of GQDs. Bending experiments indicated that the frequency drift before and after bending showed similar sensing performance with a relative error of less than 2.5%, proving the high stability of this SAW humidity sensor (Fig. 7j). In addition, the SAW humidity sensor can be placed on the human wrist to detect changes in ambient humidity, and the resonant frequency of the sensor varies readily with the on/off of the vaporizer (Fig. 7k-l).

### *B. One-Dimensional Structure*

One-dimensional graphene structures mainly refer to Graphene Nanoribbons (GNRs), which are long structures sliced from graphene lattice in one direction, with a narrow width and relatively long length. According to the different slicing directions of graphene, GNRs can be divided into four main types [228] (Fig. 8a), namely zig-zag GNRs (Fig. 8b) (i.e., the long edge of the ribbon is perpendicular to the C-C bond), chiral GNRs, armchair GNRs [229] (Fig 8c) (i.e., the long edge of the ribbon is parallel to the C-C bond), and edge-functionalized GNRs. The minimum width of GNRs can be less than 1 nm, showing significant size effects. For example, armchair graphene nanoribbons (AGNRs) typically exhibit semiconducting behavior with impressive bandgaps (2.3 eV or more). More astoundingly, the electronic structure and energy gap of AGNRs can be changed by simply adjusting the ribbon width [230]. Many meaningful works on graphene nanoribbons regarding preparation, application, and calculation [231-233] have been reported in reviews. However, there are few reports on graphene nanoribbons-based humidity sensors, one of which proposed a moisture-sensitive film based on graphene oxide nanoribbon (GOR) networks fabricated by dimensionally confined freezing self-assemble [234] (Fig. 8d). The GOR obtained through oxidation of multi-walled carbon nanotubes (MWCNTs) exhibit a homogeneous strip structure, allowing the formation of widely distributed nanopores (Fig. 8e). These nanopores can not only accelerate water trapping, but also greatly facilitate ion transportation (Fig. 8f). Results from humidity tests indicate that the resistance of GOR film decreases rapidly, with a half-saturation water adsorption period of approximately 5 ms (from 247 ms to 252 ms) as the ambient relative humidity changed from 5% to 35% (Fig. 8g), indicating that GOR film has an ultra-fast rate of hydration and excellent humidity sensitivity.

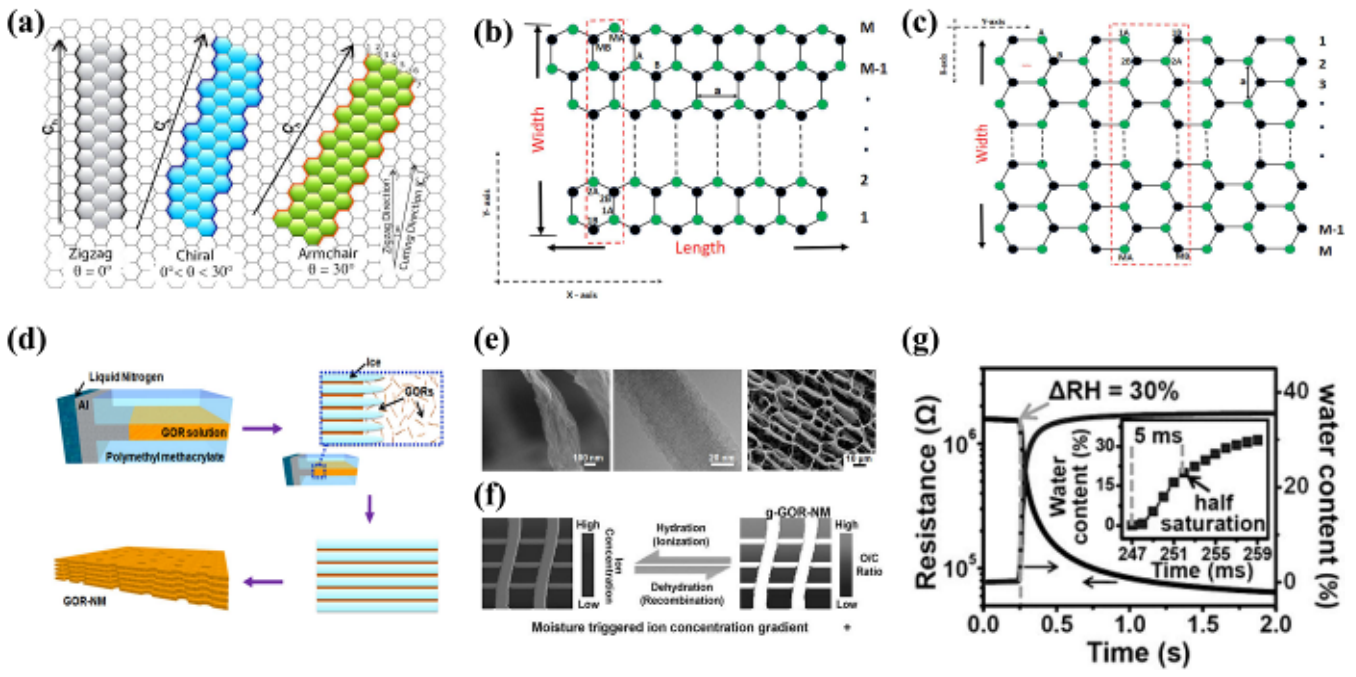

Fig. 8. One-dimensional structure of graphene-based humidity sensors. (a) Graphene nanoribbons with different structures. (b) Zig-zag GNRs (c) Armchair GNRs. (d) Preparation of moisture-sensitive thin films based on graphene oxide nanoribbon (GOR) networks. (e) Widely distributed nanopores formed by strip-structured GORs. (f) Schematic diagram of ion transport through nanopores. (g) Response/recovery time of GOR humidity sensors. Adapted from [218], [219] and [224] under a Creative Commons license.

### C. Two-Dimensional Film Structure

Few-layer 2D graphene is typically synthesized by CVD, self-assembly or chemical synthesis methods. The few-layer 2D graphene prepared by CVD method offers the advantages of controllable number of layers and higher conductivity. In 2015, Frank et al. [235] demonstrated a monolayer CVD-grown of graphene for humidity sensing. The graphene was transferred to the top surface of the Si/$SiO_2$ substrate and patterned to obtain monolayer graphene nanosheets by $O_2$ plasma etching (Fig. 9a). Monolayer graphene exhibits good moisture-sensitive and gas-sensitive properties, results show that the resistance of monolayer graphene increased as the humidity in vacuum chamber decreased. Similarly, the resistance of monolayer graphene decreased significantly with the introduction of water vapor in the test chamber (Fig. 9b). This change is closely related to the adsorption-desorption process of water molecules. In addition, the response/recovery time of the monolayer graphene sensor and commercial humidity sensors was highly consistent at 10-90% RH (Fig. 9c). Further, in 2016, Frank et al. [236] demonstrated a way for effective passivation of monolayer graphene to humidity sensing. In 2017, Frank et al. [237] demonstrated $CO_2$ sensing of monolayer graphene on the Si/$SiO_2$ substrate and analyzed its cross-sensitivity with humidity.

Experimental results indicated a significant response in the resistance of graphene when exposed to humidity. However, the electrical effects of water underneath graphene and the kinetics of how interfacial water changes with different environmental conditions have not been quantified. A metal oxide graphene device is proposed [238] for capacitive humidity sensors in response to the intercalation effect of water between graphene and $HfO_2$ (Fig. 9d). The doping effect formed by the interaction between water and oxygen will significantly change the graphene partial density of states (PDOS) as well as the relative position of the Dirac point and the Fermi level (Fig. 9e). Many advanced works have been investigated on monolayer graphene, GO and its derivatives. However, the moisture-sensitive properties of bilayer graphene have not been well studied. Fan et al. [239] prepared a bilayer graphene humidity sensor on a silicon on insulator (SOI) substrate, where the bilayer graphene was obtained through wet transfer (Fig. 9f). The resistivity of the bilayer graphene device in the vacuum chamber as a function of RH shows that a decrease in humidity (from 18% RH to 9% RH) leads to an increase in the resistance (from 477.63 Ω to 478.03 Ω) of the bilayer graphene. Another bilayer graphene device also confirmed that the resistance response decreased with the increase of relative humidity (Fig. 9g).

Both experiments and theoretical calculations revealed that the response of bilayer graphene to humidity is weaker than that of monolayer graphene. The two main types of electrical contact with graphene in the above works are bottom contact and top contact. However, one limiting factor in the development of high-performance graphene devices is the contact resistance at the metal-graphene interface, where high contact resistance can significantly affect the performance of the device. In general, graphene devices in laboratory environments may suffer from humidity variations due to the lack of encapsulation during operation. Therefore, Fan et al. [240] investigated the effect of humidity on the contact resistance at the bottom of graphene devices. Transmission line modeling (TLM) and density functional theory (DFT) simulations were applied to demonstrate that the bottom contact resistance of gold-graphene remains unaffected by humidity variations (Fig. 9h).

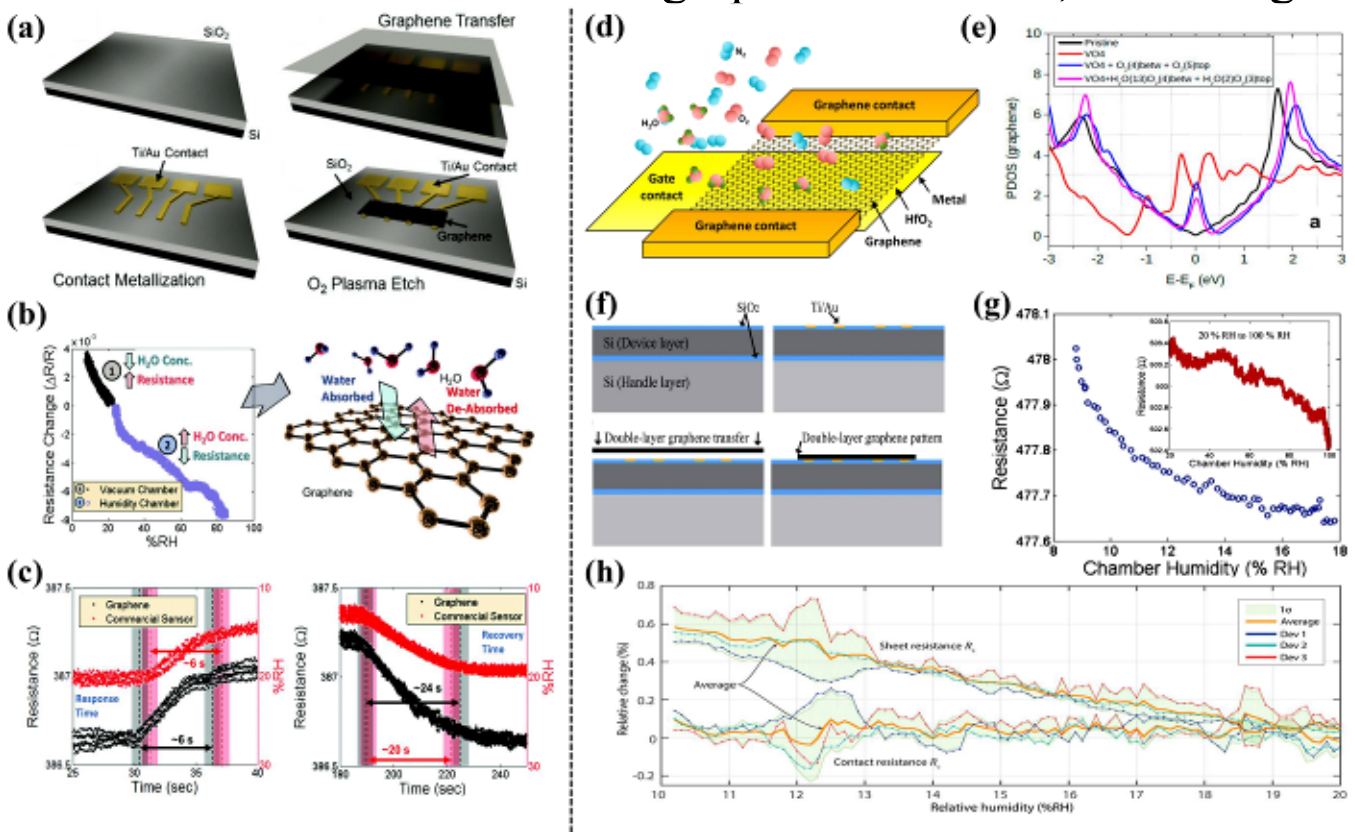

Fig. 9. Two-dimensional structures of graphene-based humidity sensors. (a) Monolayer graphene prepared based on CVD method for humidity sensing. (b) Relationship between changes in vacuum chamber humidity and resistance of monolayer graphene and schematic diagram of sensing mechanism. (c) Response/recovery time of monolayer graphene humidity sensor. (d) Schematic of a metal oxide graphene device for capacitive humidity sensing. (e) Partial density of states (PDOS) of graphene. (f) Schematic diagram of the preparation process of bilayer graphene humidity sensor. (g) Humidity response characteristics of bilayer graphene humidity sensor. (h) Effect of humidity on bottom contact resistance of graphene devices. Adapted from [225], [228] , [229] and [230] under a Creative Commons license.

Continuous exposing graphene to high humidity will lead to degradation or even failure of the sensor due to excessive water adsorption on the surface. Hence, preventing the accumulation of water to accelerate the desorption process would be the primary strategy for the preparation of high-performance humidity sensors. Zhu et al. [241] prepared humidity sensors by CVD growth of wrinkled graphene (WG) on copper-zinc (Cu-Zn) alloy (Fig. 10a). Cu-Zn alloy substrates annealed at temperatures below the melting point of Cu and above the boiling point of Zn form defects and periodic patterns that lead to the wrinkled morphology of graphene. The results showed that the rate of water evaporation from the WG surface was much higher than that of CVD graphene, and the adsorbed water could completely evaporate within 30 s. This effect can be attributed to the fact that the water droplets on the WG surface are independent, rather than forming a water film in the saturated state. SEM and AFM demonstrated that the graphene is kneaded into a graphene network ring with a diameter of ≈ 2μm (Fig. 10b-c). Furthermore, the rapid response time of the humidity sensor can be used for non-contact sensing and monitoring of respiratory rate and depth (Fig. 10d). The pre-stretching method is widely used in pressure, stress, and strain sensors due to its simplicity and maneuverability. For example, Huang et al. [242] prepared a wrinkled graphene humidity sensor for human respiration monitoring via pre-stretching and drop-coating process (Fig. 10e). SEM images showed that the rGO film exhibited a homogeneous wrinkled structure (Fig. 10f), which could inhibit water aggregation and condensation on the surface of rGO. Moreover, the wrinkled graphene humidity sensor exhibits fast response/recovery time (2.4/1.7 s) and demonstrates great potential for respiratory monitoring (Fig. 10g-h).

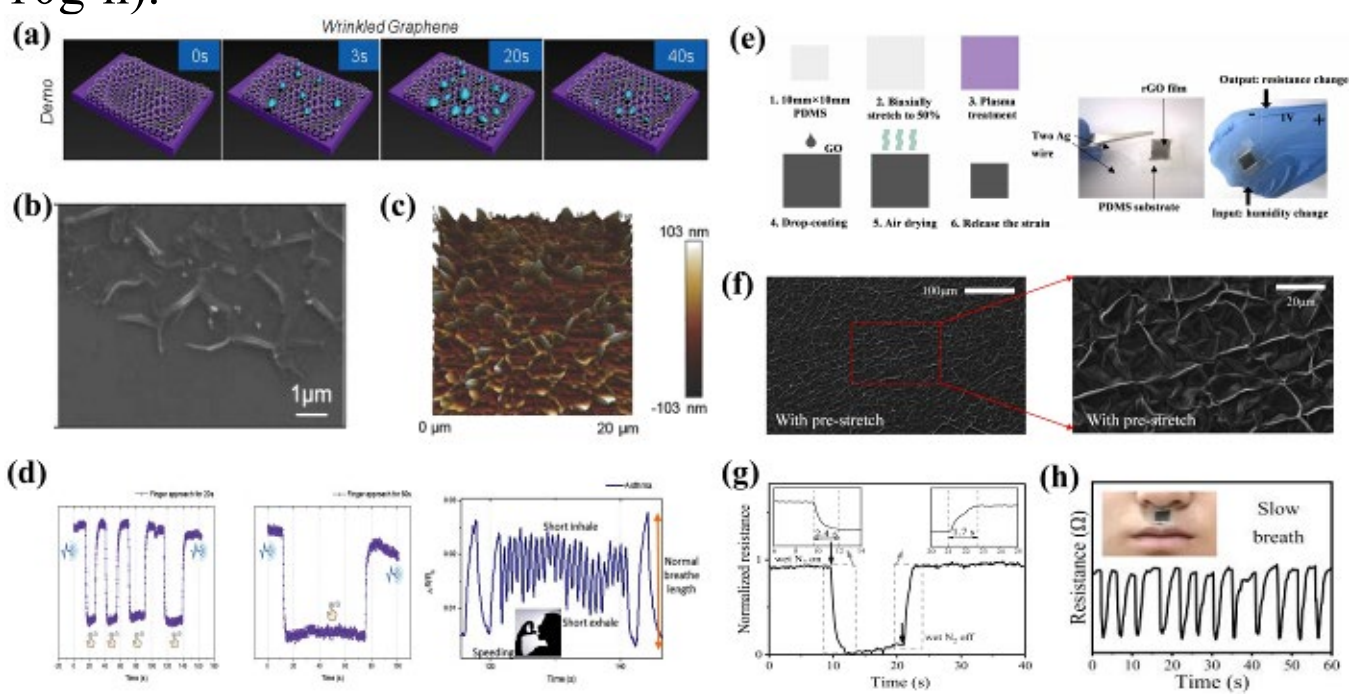

Fig. 10. Two-dimensional folded structures of graphene-based humidity sensors. (a) Schematic diagram of the rapid desorption process of wrinkled graphene (WG). (b) SEM morphology, (c) AFM morphology and (d) response properties of wrinkled graphene (WG). (e) Wrinkled graphene humidity sensor prepared by pre-stretching and drop-coating process for human respiration monitoring. (f) SEM morphology of rGO film. (g) Response/recovery time and (h) respiratory response characteristics of wrinkled graphene humidity sensors. Adapted from [231] and [232] under a Creative Commons license.

## D. Three-Dimensional Structure

Three-dimensional graphene humidity sensors exhibit significant advantages over two-dimensional graphene humidity sensors. First, the higher specific surface area of 3D graphene allows it to interact with more water molecules, improving the sensitivity of the sensor. Second, the microporous structure of 3D graphene accelerates the absorption and release of water molecules, which improves the response/recovery time of the sensor. Moreover, the flexible and stretchable properties of 3D graphene enable it to have a wide range of applications in the fields of flexible wearable electronics, motion monitoring devices, etc. Recently, Yang et al. [185] proposed a 3D structured graphene/polyvinyl alcohol/$SiO_2$ layered SAW humidity sensor prepared based on plasma jet CVD and spin coating methods. The 3D morphology of the humidity sensor consists of irregularly twisted and wrinkled 2D graphene sheets (Fig. 11a), which facilitates the adsorption and transport of water molecules. The transport effect of the 3D structured graphene and the mass loading effect of SiO2 result in fast response/recovery time (Fig. 11b), high stability, and good selectivity. Furthermore, the 3D-structured graphene humidity sensor can be used for human respiration monitoring (Fig. 11c). Furthermore, Chen et al. [243] proposed a highly sensitive microfiber humidity sensor coated with 3D graphene. The high surface area of 3D graphene greatly enhanced its interaction with water molecules, which provided the sensor with a high linear sensitivity of -4.118 dB/%RH at 79.5-85.0%RH. Recently, they presented [244] a strategy to fabricate an all-fiber humidity sensor by depositing three-dimensional graphene around the surface of free-standing microfibers (MFs). The sensitivity of this sensor was enhanced to -2.841 dB/%RH in the range of 80.3%-90.9%RH.

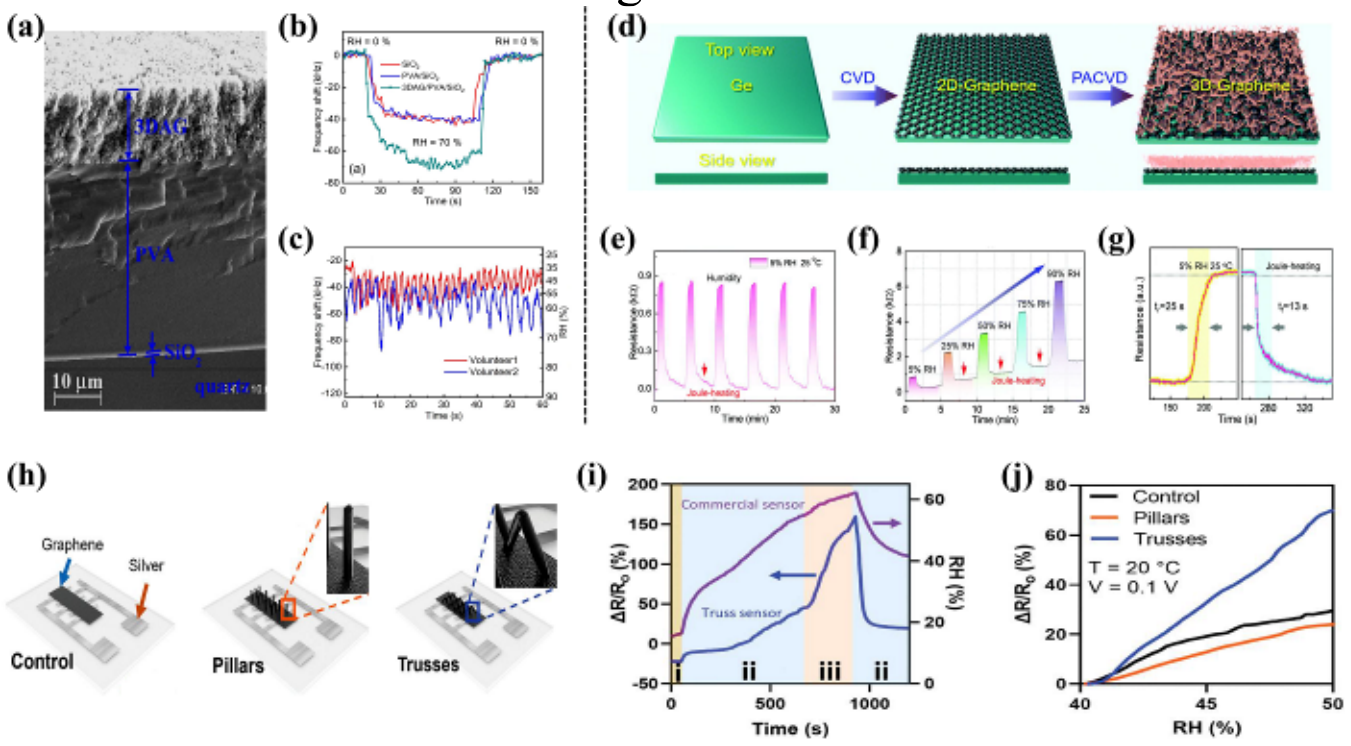

Fig. 11. Three-dimensional structure of graphene-based humidity sensors. (a) Morphology of three-dimensional structured graphene/PVA/$SiO_2$ SAW humidity sensors. (b) Response/recovery time of SAW humidity sensors. (c) Respiratory response of the SAW humidity sensor. (d) Schematic diagram of 3D graphene growth by PACVD deposition method. (e) Cyclic response curve, (f) step response curve and (g) response/recovery curve of 3D graphene

humidity sensor. (h) Schematic of aerosol jet printing to build 3D graphene humidity sensors. Performance comparison of (i) truss structure with (j) control and pillar structure humidity sensors. Adapted from [189], [235]and [236] under a Creative Commons license.

TABLE III
COMPARISON OF THE PERFORMANCE OF GRAPHENE AND ITS DERIVATIVES IN DIFFERENT DIMENSIONS FOR HUMIDITY SENSORS

| Structure type | Sensing materials | Preparation of sensors | Sensitivity/Response | Res./Rec. time | Meas. range | Ref. |
|---|---|---|---|---|---|---|
| 0D | GOQDs /CNF | Vacuum freeze drying | 51840.91 pF/%RH | 30/11 s | 11–97% | [174] |
| | GQDs | Drop coating | - | 15/55 s | 10-90% | [165] |
| | GOQDs | Drop casting | 1025.99 pF/%RH | 2.2/1 s | 22-97.3% | [176] |
| | NWs/GQDs | Dip coating | 40.16 kHz/%RH | 27/12 s | 20-90% | [184] |
| | GQDs/PEI/$SiO_2$ NPs | Drop coating | 2400 kHz/%RH | 20/5 s | 20-80% | [186] |
| | GQDs | Ink-jet printing | 7 nA/%RH | ~92/~100 s | 3.4-81.3% | [194] |
| | GQDs | Dip coating | / | 15/55 s | 40-90% | [165] |
| | GQDs | Drop casting | 390% | 12/43 s | 1-100% | [245] |
| | GQDs | Drop casting | 3.26 pm/%RH | / | 10-90% | [246] |
| 1D | GOR | Self-assembly method | / | 0.2/0.8 s | 0-90% | [234] |
| 2D | Graphene | CVD | 0.31% | 6/24 s | 10-90% | [235] |
| | Graphene | CVD | / | 5.5/8.5 s | 20-100% | [239] |
| | GO | Ink-jet printing | 72.6 nF/%RH | 2.7/4.6 s | 11-97% | [247] |
| 3D | Graphene | Dip coating | 12.927 kΩ/%RH | 1.6/2.15 s | 0-100 | [248] |
| | 3D rGO | Dip coating | -4.118 dB/%RH | 4/23.7 s | 50-85% | [243] |
| | 3D GN | Dip coating | -2.841 dB/%RH | 0.057/0.055 s | 11.6-90.9% | [244] |
| | 3D AG/PVA/$SiO_2$ | CVD | -2.429 kHz/% | 24/14.4 s | 0-90% | [185] |
| | 3D Gr | PACVD | 268.89% | 25/13 s | 5-90% | [249] |
| | 3D GF | Hydrothermal synthesis | / | 0.089/0.189 s | 0-85.9% | [250] |
| | 3D Gr | Aerosol jet Printing | 5.2 ± 2.0%Ω/%RH | / | 0-75% | [251] |

CVD methods play an important role in the preparation of graphene with various structures. For example, Wang et al. [249] prepared humidity sensors by in situ growth of 3D graphene on Ge substrate using 2D graphene as an interfacial layer via PACVD deposition method (Fig. 11d). In addition, Joule heating eliminated physically adsorbed vapor molecules from the 3D/2D-Gr hybrid structure, resulting in a fully reversible response (Fig. 11e). The hybrid structure humidity sensor exhibits excellent dynamic step response as well as outstanding response/recovery time (25/13 s) (Fig. 11f-g). The above preparation methods for 3D graphene typically rely on structural struts and redundant post-processing, greatly increasing the economic and time costs. Recently, Franklin [251] proposed a strategy to build 3D graphene microstructures using aerosol jet printing, in which 3D graphene can be printed as pillar or truss structures (Fig. 11h). The graphene truss structure adds parallel conduction paths to the sensing surface, significantly improving the sensitivity of the sensor. Moreover, the truss structure sensing humidity sensor exhibits significant performance advantages over control and pillar structure humidity sensors even in high relative humidity (Fig. 11i-j). They also demonstrated that increasing the surface area of the sensing electrodes cannot necessarily improve the sensitivity of the sensors [252, 253]. Besides, TABLE III summarizes the comparative performance of graphene and its derivatives in different dimensions for humidity sensors.

## V. GRAPHENE-BASED HUMIDITY SENSORS FOR MULTIFUNCTIONAL APPLICATIONS

Multifunctional humidity sensors have shown rapid development in various fields, and some reviews have been summarized in the areas of human respiration monitoring [22, 254, 255], plant transpiration monitoring [12], and non-contact sensing [256]. However, few reviews have reported on the construction of integrated feedback systems, human-machine interaction applications and machine learning technology of graphene-based humidity sensors. Wearable devices are usually equipped with various sensors, data processing units, Bluetooth communication technology, efficient power management systems, and feedback display systems so as to provide users with instant feedback information. In modern and future societies, human-machine interaction is a popular direction of modern information technology and artificial intelligence research, involving touch control, gesture interfaces, motion tracking, and so on.

### A. *Integrated Feedback System*

Flexible fiber products have broad application prospects in the field of wearable electronics, where e-textiles can be integrated into apparel through weaving and other processes to provide long-term, real-time strategies for human movement monitoring and health management. For example, Xu et al. [168] used GO-functionalized Coolmax fibers for a humidity sensor with excellent sensing properties, including fast response/recovery time and sensitivity. The flexible fiber sensor consists of NFC antenna system, FPGA controller, pull-up resistor, five LEDs for displaying touch actions, and power supply (Fig. 12a). However, challenges still remain to be overcome for the detection of flexible low-power humidity sensors in complex environments and the end signal processing. The humidity sensor prepared by Lee et al. [40] consists of rGO: $MoS_2$ moisture-sensitive material, finger electrodes (IDEs), temperature sensors, and miniature heaters. The back-end interface of the sensor consists of power supply, sensor unit, sensor switching unit, multiplexer (MUX), analog-to-digital converter (AD), CPU, and communication module connected with UART (Universal Asynchronous Receiver/Transmitter), USB port, status LED, DA converter, and EEPROM (Electrically Erasable Programmable read only memory). (Fig. 12b) The developed sensor system can monitor the real time variation of humidity in transformer insulating oil to predict transformer failure. Furthermore, innovative research in non-contact sensing [257-259], respiratory monitoring [260, 261] can be carried out by exploiting the sensitivity of graphene to water molecules. Integrating humidity sensors with other sensors, smart displays, and wireless Bluetooth modules can give feedback functions to sensor systems. For example, Cai et al. [262] connected the rGO/GO/rGO humidity sensor to the iPhone headphone jack and conducted humidity tests using an iOS oscilloscope app, which doubled as a signal generator and oscilloscope (Fig. 12c). Moreover, relying on big data analytics, the data collected by the humidity sensor can be uploaded to a smartphone to the cloud for in-depth analysis. Respiratory sensing systems can utilize biological synaptic structures to monitor respiratory changes, however, existing humidity sensors only respond to humidity signals and fail to achieve multimodal integrated sensing, learning as well as memory. In order to further enhance the high-precision regulation of humidity sensors, Cao et al. [263] reported an artificial respiration sensing system with high classification accuracy, which consists of GO humidity sensors, an organic electrochemical artifactual synapse (OEAS), and a microcontroller unit (MCU) module (Fig. 12d). In particular, the oxygen-rich functional groups of GO can adsorb water molecules, resulting in dramatic resistance changes of the film on interdigital electrodes (IDEs).

## B. Human Machine Interaction System

With the advent of the 5G era, human-machine interaction (HMI) smart sensors have become an important part of our daily lives and work, dramatically changing the way we communicate with our surroundings. [264-267] For example, Ren et al. [268] prepared a humidity sensor using GO as a moisture-sensitive material with LIG-IDE. In particular, the MCU records breathing signals and sends appropriate commands to control the smart car via a Bluetooth module according to the number of fast breaths. For example, one fast breathing signal, the car will move forward; two fast breaths, the car will go backward; three fast breaths, the car turns around (Fig. 12e). Furthermore, during the COVID-19 epidemic, the contact action between the human body and switches inevitably leads to mechanical wear and tear as well as bacterial or viral cross-infection between users. [269-271]

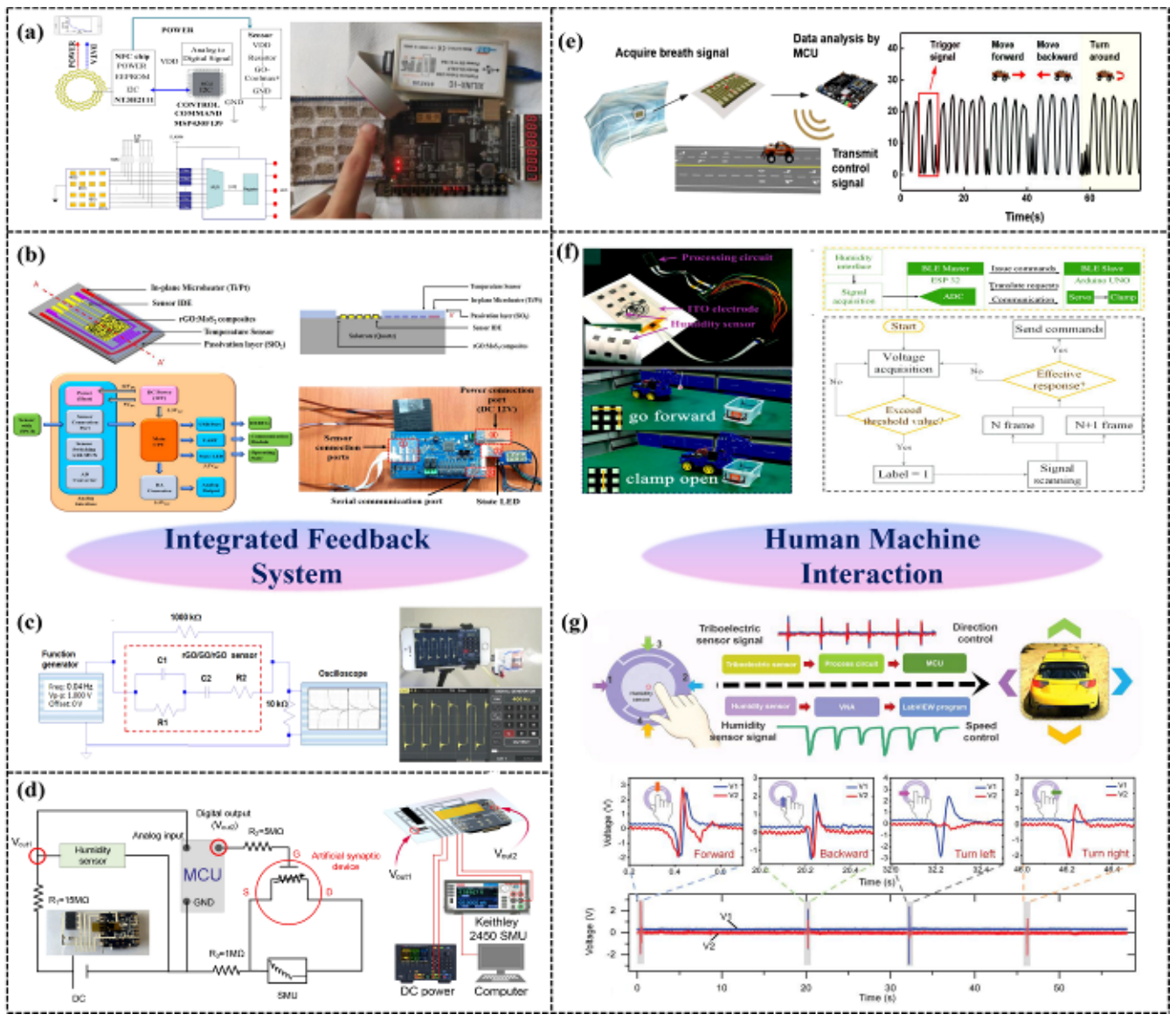

Fig. 12. Integrated feedback and human machine interaction system based on graphene humidity sensors. (a) Circuit diagram and image of Coolmax fabric sensor woven into 4 × 4 keyboard. (b) Schematic structure of rGO/$MoS_2$ based humidity sensor and its interface design. (c) Diagram of wireless transmission measurement circuit system based on rGO/GO/rGO humidity sensor. (d) Diagram of circuit connection of artificial respiratory perception system. (e) Schematic of a non-contact human-machine interaction (HMI) device and the process of breathing to control a smart car. (f) Schematic diagram of sensor array, robotic car and flowchart of motion recognition algorithm for trolley operation in non-contact HMI system. (g) Process for non-contact vehicle control with multimodal sensors and two-channel triboelectric voltage generated by finger motion. Adapted from [172], [40] , [252] , [253] , [258] , [153] and [262] under a Creative Commons license.

Therefore, HMI is highly necessary in non-contact control systems. With the help of motion recognition algorithms, the successful preparation of humidity sensors based on the combination of electrostatic spinning and graphene realizes the non-contact control of the robot to monitor the delivery of pills. [156] (Fig. 12f) The accurate recognition of non-contact sliding movements of the fingers is of great significance for the realization of human-robot interaction, and provides an effective strategy for the development of intelligent electronic devices that require non-contact scenarios. In order to verify the feasibility of human-machine interaction in practical applications, Lee et al. [272] designed and integrated a humidity and triboelectric dual-sensor device based on GO film coating. In particular, the humidity sensor and the triboelectric sensor are used to recognize command signals and transient signals during non-contact interaction, respectively. The outstanding performance of the sensors can be successfully applied to VR vehicle control through intuitive finger interaction (Fig. 12g).

## C. Machine Learning for Humidity Sensors

Machine learning and artificial intelligence techniques can be combined with sensor data fusion to provide in-depth analysis of data collected by humidity sensors through algorithms. Machine learning algorithms can be used to realize functions such as smoothing humidity data, detecting outliers and predicting future humidity changes. With the development of wearable electronic systems, humidity sensors are beginning to be used for human body related (HBR) humidity detection. Utilizing machine learning algorithms can provide more accurate health monitoring and disease prevention. For example, Zhou et al. [273] prepared a conductive ink (GC) using graphene nanosheets and carbon nanotubes. In this, the GC ink was used as a fusion agent (FA) for multi-jet fusion (MJF) printing, and thermoplastic polyurethane (TPU) powder was used to produce an adaptable and pliable polymer layer, which in turn provided deformation (Fig. 13a). The prepared multilayer GC sensors can be used to predict human movement and health conditions. Statistical features of the data collected from the GC humidity sensor were used to train a support vector machine (SVM) to predict the breathing pattern of a particular individual. The trained model achieves 100% classification accuracy by both linear and polynomial transformations (Fig. 13b).

By combining different types of sensors (e.g., humidity, temperature, haptics, gases, etc.) to collect data, multimodal sensing technology can provide more comprehensive environmental sensing than a single sensor. This multidimensional sensing capability allows systems to understand and interpret their surroundings more accurately and thus make more precise decisions. Bai et al. [274] proposed a multimodal respiratory sensing system that integrates an ammonia sensor based on thermally cleaved conjugated polymers, a rGO-based humidity sensor, and a respiratory dynamics sensor. The multimodal signals collected from the respiratory sensors are decoded using a machine learning algorithm with K-mean clustering to discriminate between ammonia concentration and humidity levels in healthy and diseased respiration (Fig. 13c).

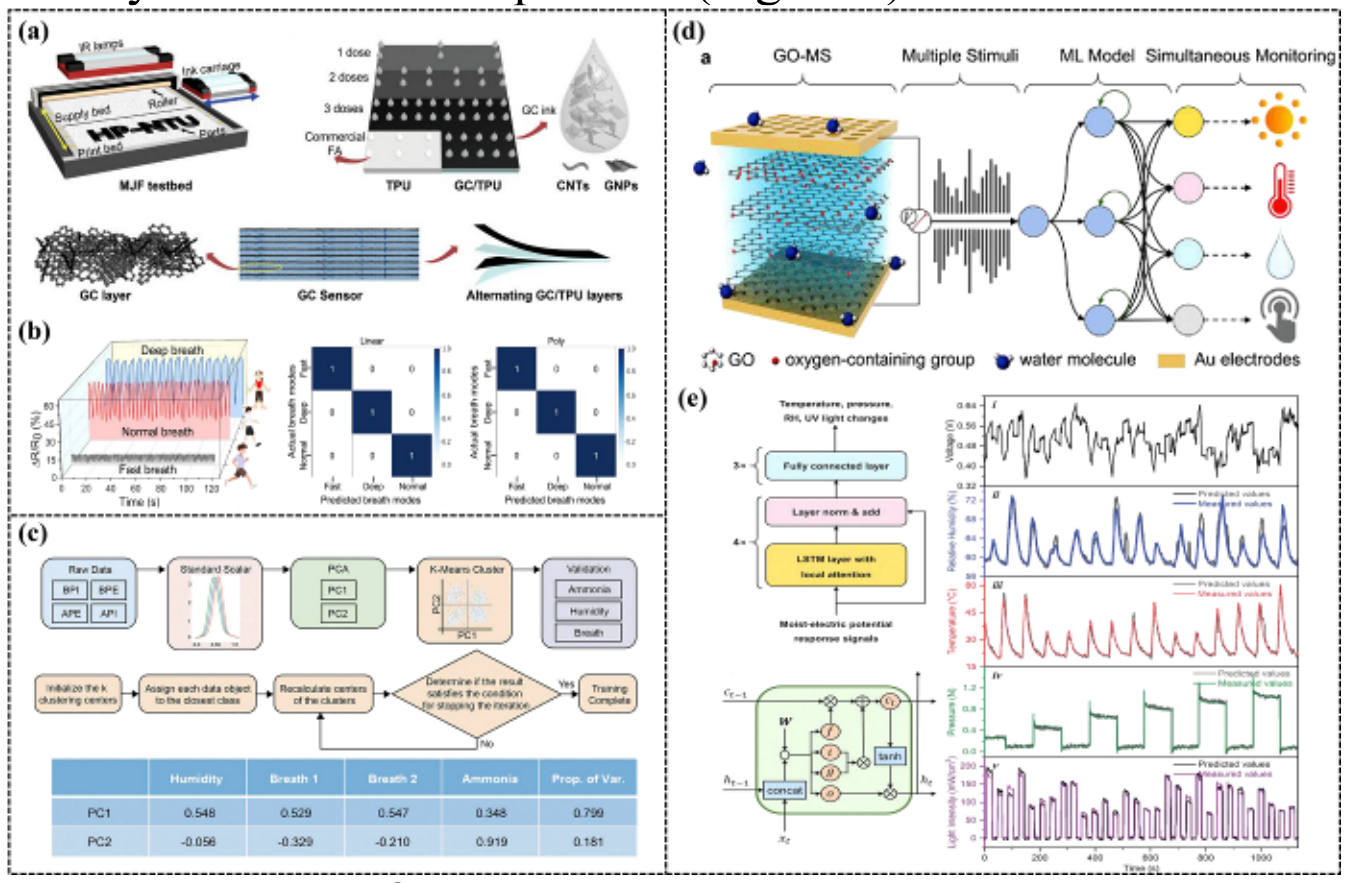

Fig. 13. Integrated feedback and human machine interaction system based on graphene humidity sensors. (a) MJF printing technology. (b) The result of training SVM to predict the breathing pattern of a specific individual. (c) Block diagram of machine learning algorithm and K-means clustering steps for decoupling signals from humidity sensors, respiration sensors and ammonia sensors. (d) Schematic diagram of moist-electric GO-MS for simultaneous multiple stimuli monitoring. (e) Multi-stimulus perception and decoupling analysis. Adapted from [265], [266] and [267] under a Creative Commons license.

Presently, it is difficult for existing multifunctional material molding monitoring systems to achieve simultaneous response to multiple stimuli and predictable signal decoupling, especially independent energy supply. To address this issue, Qu et al. [275] developed a GO single-component multimodal sensor (GO-MS) prepared from GO/sodium alginate by freeze-drying combined with directed thermal reduction (Fig. 13d). The sensor is capable of simultaneously monitoring multiple environmental stimuli through a self-powered unit with unique humidity and generates a sustainable electrical potential by spontaneously adsorbing water molecules in the air. Furthermore, the machine learning (ML) decoupling model consists of four long short-term memory (LSTM) layers with local attention mechanisms and three fully connected layers. Based on this well-designed decoupled ML model, the GO-MS is able to simultaneously monitor and decouple real-time changes in the four stimuli (Fig. 13e).

## VI. CONCLUDING REMARKS AND PERSPECTIVES

In the last decade, humidity sensors based on novel materials such as optical fibers, nanomaterials, cellulose, MXene, etc. have been profoundly outlooked and summarized in the fields of non-contact measurements, human health care, plant transpiration, and human-computer interaction. With the further research on graphene materials, graphene-based humidity sensors will be more widely used in different fields in the future. According to this review, rational structural design is beneficial to enhance the miniaturization, portability, and sensitivity of the device. Various graphene preparation methods and structural design provide a new path to develop high-quality and low-cost graphene products, significantly expanding their applications in energy, environment, and future wearable electronic devices. Meanwhile, special attention needs to be paid to the following directions of development:

(1) The synergistic effect of high-performance humidity sensing materials with different structures of graphene needs to be further explored, especially the emerging two-dimensional materials in recent years. The development of new heterostructures can further promote the development of graphene humidity sensors.

(2) The reasonable customization of graphene structures according to experiments and market demands is of great significance.

(3) The proper use of improved pattern recognition algorithms, 5G wireless networks, and machine learning are important strategies to fulfill the growing demand for smart and wearable electronic devices.

(4) Future interdisciplinary researchers should work closely to bridge the practical gap between graphene humidity sensors in laboratory and practical applications. We sincerely hope that this critical review will provide a comprehensive understanding of graphene humidity sensors and guide their development towards practicality and reliability.

However, several challenges still exist for graphene humidity sensors in performance research and practical applications:

(1) The quality of graphene obtained from different preparation methods is uneven.

(2) The problems of pollution, energy consumption, as well as time cost of graphene preparation process pose great challenge for large-scale production. An effective strategy is to optimize the parameters and preparation process according to the market demand to maximize the production efficiency.

(3) The sensing mechanism of graphene humidity sensors is still unclear. It is commonly believed that the interaction between graphene and adsorbed molecules is physical and chemical adsorption, while quantitative analysis on the microscopic scale (e.g., electron flow, molecular diffusion) is not clear. Hence, rational combination of solid-state physics, quantum mechanics, and surface chemistry can realize a systematic interpretation of the sensing process and further deepen the understanding of the graphene interaction with adsorbed molecules.

(4) The performance of graphene humidity sensors needs to be further improved. The presently proposed graphene humidity sensors are difficult to operate continuously at high humidity (> 90%), which in turn leads to a significant decline in sensitivity, selectivity, response/recovery ability, stability, and repeatability of the sensors. An effective strategy is to develop and design graphene materials with different structures (0D, 1D, 2D, 3D structures) to synergize with other materials.

In the future, with the development of microelectronics technology, the size of humidity sensors will become smaller and smaller, which enables them to be integrated into smaller devices, such as smartphones and smartwatches. Additionally, the development of artificial intelligence technology will make humidity sensors more intelligent. For example, through machine learning technology, humidity sensors can automatically recognize changes in temperature and humidity in the environment, and automatically adjust the working state of the device to improve the operational efficiency of the device. At the same time, advances in wireless communication technology and sensor fusion technology will enable humidity sensors to realize wireless connectivity, making it easier for sensors to be deployed and monitored, while reducing wiring costs and maintenance costs.

## Acknowledgment

Hongliang Ma, Jie Ding, Zhe Zhang, Qiang Gao, Quan Liu and Gaohan Wang are with the Advanced Research Institute for Multidisciplinary Science, Beijing Institute of Technology, Beijing 100081, China, also with the School of Integrated Circuits and Electronics, Beijing Institute of Technology, Beijing 100081, China.

Wendong Zhang is with the State Key Laboratory of Dynamic Measurement Technology, North University of China, Taiyuan 030051, China, also with the National Key Laboratory for Electronic Measurement Technology, School of Instrument and Electronics, North University of China, Taiyuan 030051, China. (e-mail: wdzhang@nuc.edu.cn)

Xuge Fan is with the Advanced Research Institute for Multidisciplinary Science, Beijing Institute of Technology, Beijing 100081, China, also with the School of Integrated Circuits and Electronics, Beijing Institute of Technology, Beijing 100081, China. (e-mail: xgfan@bit.edu.cn).

(Corresponding authors: Xuge Fan and Wendong Zhang)

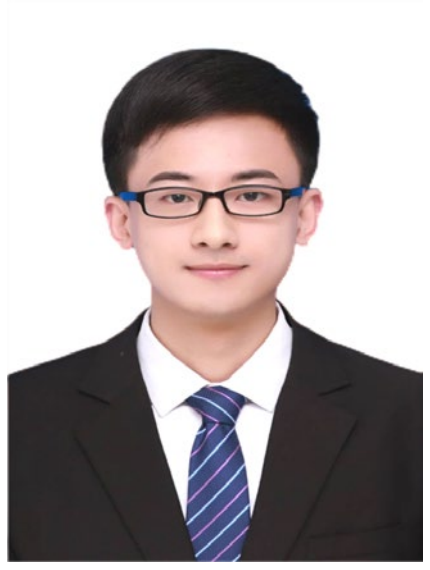

**Hongliang Ma** received his master’s degree from Qilu University of Technology in 2020. He is currently a Ph. D. candidate at the School of Integrated Circuits and Electronics, Beijing Institute of Technology at Beijing, China, under the guidance of Prof. Xuge Fan.

His current research interests include the application of 2-dimensional material electronics in humidity and gas sensing.

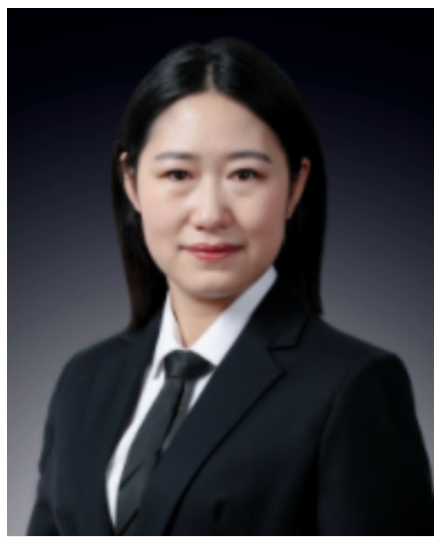

**Jie Ding** received the Ph.D. degrees from the University of Glasgow, UK in 2015.

She is now a pre-appointed associate professor at Beijing Institute of Technology.

Her current research interests include Modeling and Simulation of Novel Semiconductor Devices, Random Rise and Fall and Reliability Studies of CMOS Devices and Flash Memory Devices, Design and Process Co-Optimization Methodology (DTCO), Modeling and Simulation of Sensor Devices and Mechanism Studies.

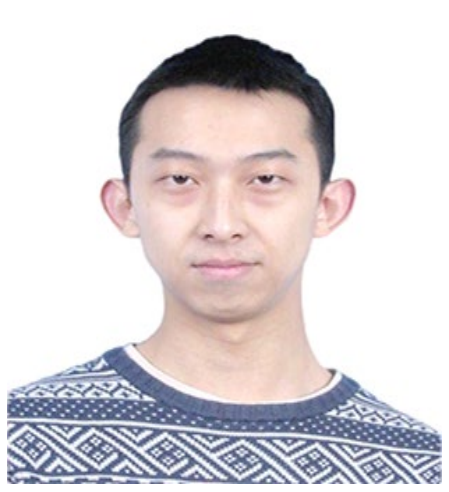

**Zhe Zhang** received his bachelor's degree from Taiyuan University of Technology in 2021. He is currently pursuing his master's degree in the Department of Electronic Information at Beijing Institute of Technology under the supervision of Prof. Xuge Fan.

His current research interests include micro-nano electromechanical sensors based on 2D materials such as graphene.

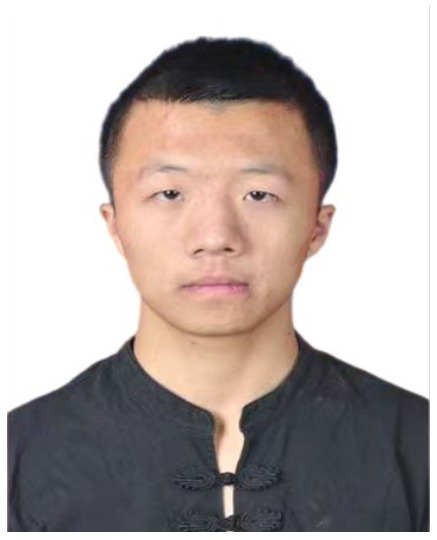

**Qiang Gao** received his bachelor's degree from Zhengzhou University of Aeronautic in 2022. He is currently studying for a master's degree in Fan Xuge's research group at Beijing Institute of Technology.

His current research interests include graphene acceleration sensors, graphene humidity sensors, etc.

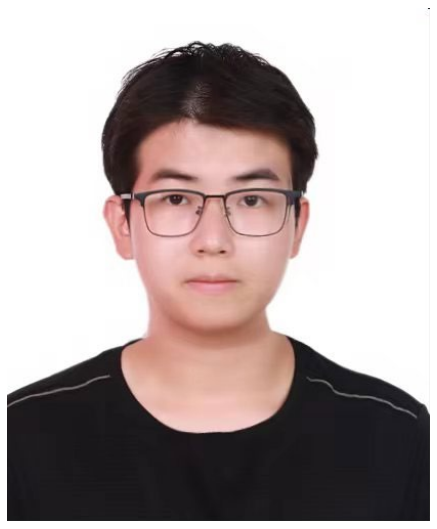

**Quan Liu** received his bachelor's degree from Taiyuan University of Technology in 2019. He is currently pursuing his master's degree at Beijing Institute of Technology under the supervision of Prof. Xuge Fan.

His current research interests include two-dimensional materials such as graphene and their sensing applications.

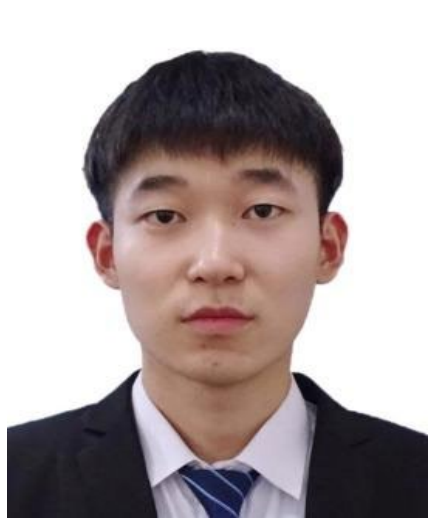

**Gaohan Wang** received his bachelor's degree from Taiyuan University of Technology in 2019.He is currently pursuing a master's degree in the School of Integrated Circuit and Electronics at Beijing Institute of Technology, under the guidance of Professor Xuge Fan.

His current research interests include graphene and other two-dimensional materials and their applications in sensors.

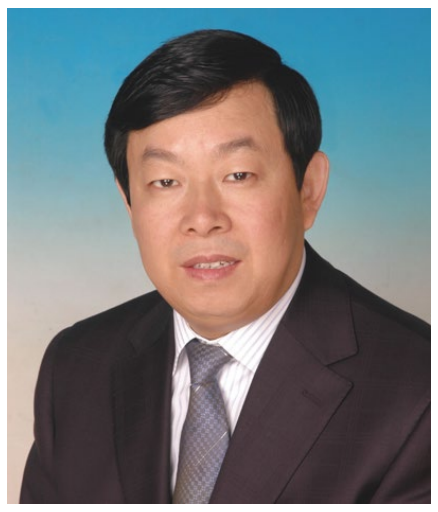

**Wendong Zhang** received the Ph.D. degrees in 1995 from Beijing University of technology. From 1996 to 1998, and worked at the Postdoctoral Station of Instrumentation at Tsinghua University from 1996 to 1998, and served as a Senior Visiting Scholar at the Center for Sensors and Actuators of the University of California, Berkeley, USA, and the Research Institute of Industrial Technology of the Ministry of International Trade and Industry (MITI), Japan, in 1998 and 2001, respectively.

His current research interests include dynamic test technology and intelligent instrumentation, micro-electromechanical systems (MEMS) and other areas of research.

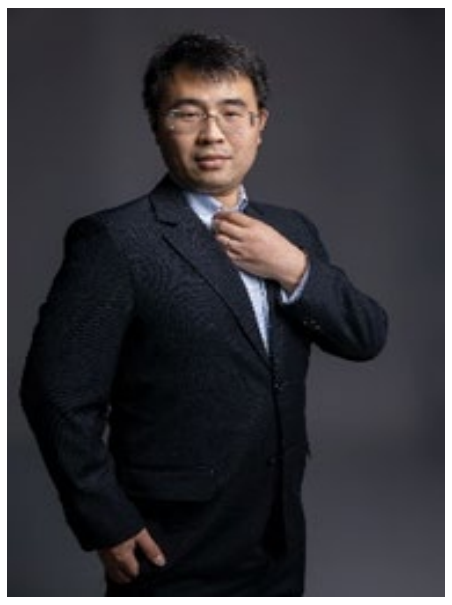

**Xuge Fan** received the M.S. degrees from the School of Taiyuan University of Technology in 2013 and Ph.D. degrees from the School of Royal Institute of Technology (KTH), Sweden in 2018. He's now a professor at Beijing Institute of Technology.

His current research interests include Micro-Nano Electro-Mechanical Systems (MEMS/NEMS), 2D materials such as graphene, micro-nano sensors, etc.